\documentclass[aps,twocolumn,prb,amsmath,amssymb,floatfix,reprint]{revtex4-2} 

\usepackage[utf8]{inputenc}
\usepackage{graphicx}
\usepackage{bm}
\usepackage{color}
\usepackage[colorlinks=true,citecolor=blue]{hyperref}
\usepackage{graphicx}
\usepackage{braket}
\usepackage{physics} 
\usepackage{comment} 
\makeatletter
\AtBeginDocument{\let\LS@rot\@undefined}
\makeatother

\begin{document}

\title{Entanglement dynamics in minimal Kitaev chains}
\author{Vimalesh Kumar Vimal}

\author{Jorge Cayao}
%\email[]{jorge.cayao@physics.uu.se}
\affiliation{Department of Physics and Astronomy, Uppsala University, Box 516, S-751 20 Uppsala, Sweden}
\date{\today}

\begin{abstract}
Minimal Kitaev chains host Majorana quasiparticles, which, although   not topologically protected,  exhibit spatial nonlocality and hence expected to be useful for quantum information tasks. In this work, we consider two- and three-site Kitaev chains and investigate the   dynamics of bipartite and multipartite entanglement by means of   concurrence and   geometric measure of entanglement. In two-site Kitaev chains, we find that maximally entangled states can robustly emerge, with their stability and periodicity highly controllable by the interplay between the superconducting pair potential and the onsite energies. At the finely tuned sweet spot, where Majorana quasiparticles appear, the system exhibits oscillations between separable and entangled states, whereas detuning introduces tunable valleys in the entanglement dynamics. Extending to the three-site Kitaev chain, we uncover   rich bipartite and multipartite entanglement  by generalizing the concepts of concurrence and geometric measure of entanglement. At the  sweet spot, the  Majorana   quasiparticles emerging at the edges suppress concurrence between the edges, while a finite detuning is able to restore it. Depending on the initial state, the three-site Kitaev chain can dynamically generate   either a maximally    entangled Greenberger–Horne–Zeilinger state or an imperfect W-type state exhibiting multipartite entanglement, although a (maximally entangled) pure  W state cannot be realised due to parity constraints.  Our results provide a resource for generating and characterising highly entangled states in minimal Kitaev chains, with potential relevance for  quantum applications.
\end{abstract}
\maketitle
 
 %%%%%%%%%%%%%%%%
%     Section I:  Introduction     %
%%%%%%%%%%%%%%%%

\section{Introduction}
Topological superconductors host exotic quasiparticles known as Majorana zero modes (MZMs) and have attracted considerable interest in the past decade \cite{tanaka2011odd_review,doi:10.7566/JPSJ.85.072001,Sato_2017,Aguadoreview17,lutchyn2018majorana,Cayao2020odd,frolov2019quest,flensberg2021engineered,10.1093/ptep/ptae065,Fukaya_2025}  due to their potential applications in fault-tolerant quantum computation~\cite{sarma2015majorana,Lahtinen_2017, beenakker2019search, aguado2020majorana, Marra_2022, qx36-4rv1}.   MZMs were initially predicted to emerge in the topological phase of the Kitaev chain \cite{kitaev2001,10.1093/ptep/ptae065}, which consists of  spinless fermions  with $p$-wave superconductivity and can be effectively realized in semiconductor-superconductor hybrid systems \cite{PhysRevLett.103.020401, PhysRevB.82.134521, Lutchyn2010, Oreg2010}. MZMs appear as self-conjugate zero-energy states located at the edges of the topological superconductor \cite{10.1093/ptep/ptae065}, thereby exhibiting a unique  spatial nonlocality  that makes them promising for designing topological qubits \cite{aguado2020perspective,aguado2020majorana} and for storing information in a nonlocal manner \cite{sarma2015majorana}. While the nonlocal properties of MZMs are certainly promising \cite{PhysRevB.96.085418,PhysRevB.96.201109,PhysRevB.96.205425,PhysRevB.104.L020501,PhysRevB.109.L081405,PhysRevB.110.224510,79tj-c3y4,8mzs-dx7h}, their experimental realization is still unclear due to the lack of unequivocal  evidence for the topological phase \cite{cayao2015,prada2019andreev,valentini2021nontopological,Aghaee2025}.

In an alternative approach, Majorana quasiparticles have been shown to emerge in minimal two-site Kitaev chains  \cite{Leijnse2012,Sau2012,PhysRevB.106.L201404,PRXQuantum.5.010323,PhysRevResearch.5.043182,Liu2024,Alvarado_2024,r75t-jv32,PhysRevB.109.035415,PhysRevB.110.125408,Cayao_2025,PhysRevB.111.115419,PhysRevB.110.245412,cayao2025nonLocalDiode}, which can be realized by coupling two quantum dots (QD) via a superconductor-semiconductor system \cite{AguadoSouto2024}. In this setup,  Majorana quasiparticles appear in each site or QD when electron cotunneling and crossed Andreev reflection are equal \cite{Leijnse2012}, a fine-tuned point referred to as sweet spot that does not involve any topological protection. For this reason, 
 Majorana quasiparticles in a minimal Kitaev chain are often referred to as  poor man’s Majorana modes (PMMMs) and there already exist convincing experimental evidence of their realization in superconductor-semiconductor systems \cite{Dvir2023,bordin2023tunable,bordin2024crossed,zatelli2024robust,ten_Haaf2024,van_Loo2026}.
 
Building on these ideas, three-site Kitaev chains have been theoretically proposed \cite{PhysRevB.111.115419,PhysRevB.111.235409,2r8x-9d9m} and experimentally fabricated \cite{ten_Haaf2025,Bordin2025,Bordin2026},  demonstrating the high controllability of the QD-based systems for exploring PMMMs and their signatures. Despite their lack of topological protection, PMMMs  exhibit key Majorana signatures, such as spatial nonlocality and nontrivial coherence \cite{Leijnse2012,PhysRevB.110.125408}, making PMMMs in minimal Kitaev chains to be of an intrinsic relevance for several Majorana-like uses \cite{AguadoSouto2024}. This intrinsic nonlocality raises the natural question of how entanglement develops and evolves in minimal Kitaev chains, thereby highlighting their potential for practical quantum information applications~\cite{PhysRevB.110.224510,dxb2-15jf, PhysRevB.111.075415, Chen_2018}. In this regard,  central has been the interest in the entanglement generation and the realization of highly entangled states in systems formed by QDs and Majorana states \cite{PhysRevB.87.214513,PhysRevB.91.214507,Shi_2016}; see also \footnote{ {\color{red}It is worth noting that entanglement has also been widely used as probe of topological phase transitions in several large Kitaev systems via the entanglement entropy and spectrum, see e. g., Refs.\,\cite{cv5q-8t25,PRXQuantum.5.010313,PhysRevResearch.2.013175,PhysRevB.96.115108, universe5010033, PhysRevLett.119.250401}}}. The focus in these studies was on nonlocal correlations \cite{PhysRevB.87.214513,PhysRevB.91.214507}, parity effects \cite{PhysRevB.91.214507,Shi_2016}, entanglement dynamics \cite{Shi_2016,Xu_2012}, as well as on the role of environmental coupling and decoherence on quantum correlations \cite{PhysRevA.80.062322,delValle:11,PhysRevA.99.042320,Xu_2012,KENFACK2017123}. Despite these interesting advances,  bipartite entanglement and multipartite quantum correlations in  minimal Kitaev chains has remained seldom explored.

\begin{figure}[!t]
\centering
\includegraphics[width =0.8\linewidth]{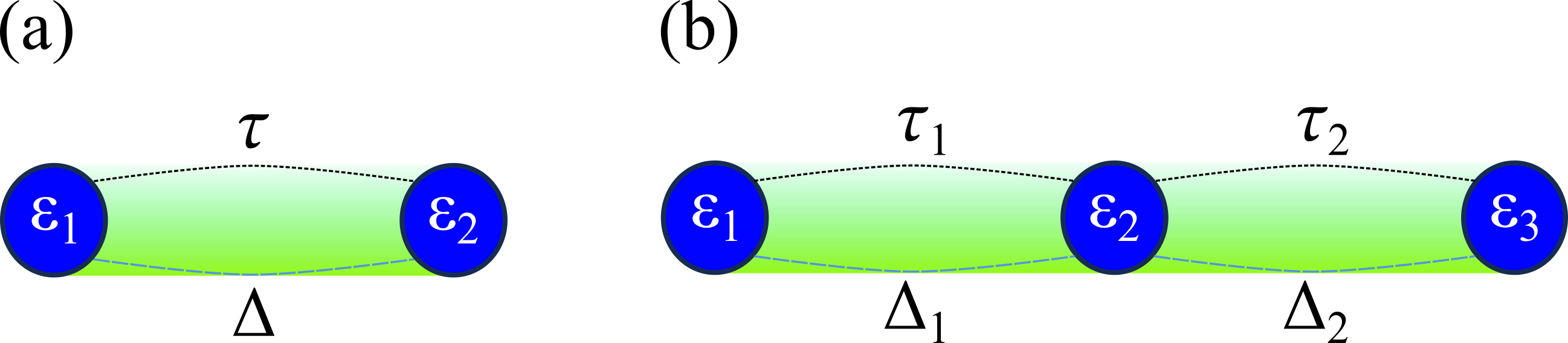}
\caption{Illustration of minimal Kitaev chains realised in QDs (blue) coupled by superconductor-semiconductor systems (light green). (a) A two-site Kitaev chain: QDs  with onsite energies $\varepsilon_i$  are connected via a superconductor-semiconductor segment, which induces a single-electron tunnelling  hopping amplitude ($\tau$) and a spin-polarized $p$-wave pair potential ($\Delta$). (b) A three-site Kitaev chain: QDs with onsite energies $\varepsilon_i$ are coupled through a superconductor-semiconductor hybrid, giving rise to nearest-neighbour hopping amplitudes $\tau_{1,2}$ and $p$-wave pair potentials $\Delta_{1,2}$.}
\label{sketch}
\end{figure}
 
Motivated by the recent progress on minimal Kitaev chains, in this work we investigate the entanglement dynamics in two- and three-site Kitaev chains, as illustrated in Fig.~\ref{sketch}. In particular, we quantify the dynamics of bipartite~\cite{Wootters2001, PhysRevA.98.052303, Osterloh2002} and multipartite entanglement~\cite{PhysRevA.68.042307, PhysRevA.62.062314, PhysRevA.76.012326, MA_2011, Ma_Xiao-San_2010, PhysRevLett.101.025701, PhysRevA.105.062202, Xiong2022, Gulati2025} using measures such as concurrence and geometric measure of entanglement (GME) \cite{PhysRevA.68.042307, PhysRevLett.101.025701, Xiong2022, PhysRevA.80.032324, PhysRevA.93.012304, Streltsov_2010, Shimony1995, PhysRevA.80.032324, QI20071363, Ramachandran2025, MISHRA2024100230, Weinbrenner_2025}, which we also support  by  analysing the  return probability and entanglement dynamics.  We demonstrate that, in the two-site Kitaev chain, maximally entangled states can be dynamically generated from any initial state in either parity sector, with the peculiarity that the stability and periodicity of entanglement characterise distinct parameter regimes. At the sweet spot, the system oscillates between separable and  maximally entangled states. Detuning the onsite potentials, which involves the leakage of the wavefunctions  of  PMMMs into each QD,  introduces entanglement valleys between successive maximally entangled states, whose  depth and width   exhibit a high  controllability by the onsite energies.

In the case of a three-site Kitaev chain and using  the measures discussed above, we uncover a richer entanglement dynamics, including bipartite and multipartite entanglement. At the genuine sweet spot~\cite{PhysRevB.111.235409},  we show that entanglement between the outer QDs vanishes due to the localisation of PMMMs, while the nearest-neighbour entanglement reaches only half of its maximum value. We also obtain that a finite detuning from the sweet spot enhances   entanglement not only  between nearest QDs but also between the outer QDs, making the three-site Kitaev setup of interest even without hosting PMMMs. 

Moreover, we unveil that, depending on the parity of the initial state, the three-site Kitaev chain   can dynamically generate multipartite entanglement of either Greenberger-Horne-Zeilinger (GHZ)- \cite{Greenberger1989,10.1119/1.16243} or imperfect W-type state,  although a perfect W state \cite{PhysRevA.62.062314, PhysRevA.68.042307} cannot be achieved within this Kitaev chain. Both GHZ \cite{Greenberger1989,10.1119/1.16243}  and W \cite{PhysRevA.62.062314, PhysRevA.68.042307} states are entangled states of three qubits that offer complementary types of entanglement due to their key role  for an essential three-way and pairwise entanglement \cite{DHondt2006}, respectively. These states are relevant for  quantum metrology \cite{giovannetti2011advances,PhysRevLett.134.130604} and  various quantum information tasks \cite{pirandola2017fundamental,PhysRevA.98.052320,PRXQuantum.4.040323}, and their generation and characterization have been explored in distinct platforms  \cite{doi:10.1021/acs.nanolett.3c01551, Yang:2024jug, PhysRevA.91.022312, Resch2005, Mikami2005, Nelson2000, Laflamme1998,bao2024creating}. Our findings therefore provide a systematic characterization of entanglement dynamics in minimal  Kitaev chains and demonstrate the generation of useful  highly entangled states. This, in turn, unveils  the   potential of minimal Kitaev chain chains as   relevant platforms for controllable bipartite and genuine multipartite entanglement generation.

This paper is organised as follows. In Sec.~\ref{SectionII}, we introduce the Hamiltonians of the two- and three-site Kitaev chains, analyse their eigenenergies and eigenstates, and define the relevant parameter regimes with PMMMs. In Sec.~\ref{SectionIII}, we examine the quantum dynamics of the two-site Kitaev chain, focusing on various entanglement measures and dynamical observables for different initial states and parameter regimes. In Sec.~\ref{SectionIV}, we extend this analysis to a three-site Kitaev chain, computing the same observables both at and away from the sweet spot. Sec.~\ref{section5} is devoted to the study of multipartite entanglement in the three-site chain, where we characterise the emergence of GHZ- and W-type states in different regimes.  In Sec.~\ref{SectionVI}, we summarise our main results and discuss their implications. For completeness, the dynamics of the odd-parity sector of the three-site Kitaev chain are presented in Appendix~\ref{oddDyn3sites}.

%%%%%%%%%%%%%%
%     Section II:  Model     %
%%%%%%%%%%%%%%

\section{Models  for two- and three-site Kitaev chains}
\label{SectionII}
We start by introducing the models of two- and three-site Kitaev chains. A two-site Kitaev chain is modelled by the following Hamiltonian
\begin{equation}
\label{Hamiltonian23a}
H_{2}=\sum_{i=1}^{2}\varepsilon_i\,c_i^\dagger c_i 
       +  \bigl(\tau\,c_1^\dagger c_2 
       + \Delta\,c_1^\dagger c_2^\dagger + \text{h.c.}\bigr)\,,
\end{equation}
while a three-site Kitaev chain by
\begin{equation}
\label{Hamiltonian23b}
H_{3} = \sum_{i=1}^{3}\varepsilon_i\,c_i^\dagger c_i 
       +  \sum_{i=1}^{2}\bigl(\tau_i\,c_i^\dagger c_{i+1} 
       + \Delta_i\,c_i^\dagger c_{i+1}^\dagger + \text{h.c.}\bigr)\,.
\end{equation}
 Here, $c_i^\dagger$ ($c_i$) creates (annihilates) an electronic state at the  $i$-th  site (or QD)   and $\varepsilon_i$ is the on‐site   energy. In the two-site Kitaev chain [Eq.\,(\ref{Hamiltonian23a})], $\tau$ denotes the hopping amplitude characterizing ECT, while $\Delta$ represents the $p$-wave pair potential due to CAR. In the three-site Kitaev chain, $\tau_{i}$ is the nearest-neighbour hopping amplitude, while $\Delta_i$ the $p$-wave pair potential between sites. Both systems can host PMMMs \cite{Leijnse2012,Cayao_2025,PhysRevB.110.125408,PhysRevB.111.115419,PhysRevB.111.235409,2r8x-9d9m,AguadoSouto2024}   and experimental evidence for their realisation has already been reported in superconductor–semiconductor hybrids connected by QDs, see Refs.\,\cite{Dvir2023,bordin2023tunable,bordin2024crossed,zatelli2024robust,ten_Haaf2024,van_Loo2026} and \cite{ten_Haaf2025,Bordin2025,Bordin2026}. Without loss of generality,  we refer to the sites as QDs throughout this work where appropriate. In the two-QD Kitaev chain, PMMMs emerge when the parameters are fine-tuned to $\varepsilon_i = 0$ and $\tau = \Delta$: Under these conditions,   PMMMs become localised in each QD   and are quadratically protected against perturbations in the onsite energies~\cite{Leijnse2012}. More precisely, at the sweet spot, we can write   Eq.\,(\ref{Hamiltonian23a}) in the  Nambu   basis $\psi = (c_1,\,c_2,\,c_1^\dagger,\,c_2^\dagger)$ and directly verify that the zero energy solutions lead to two PMMMs ($\gamma_{1,2}$) located in each site \cite{Leijnse2012}, namely,   $\gamma_{1} = (c_1 + c_1^\dagger)/\sqrt{2}$ and $\gamma_{2} = i\,(c_2 - c_2^\dagger)/\sqrt{2}$. In a three-QD Kitaev chain, the sweet spot is achieved at  $\varepsilon_i=0$ and $\tau_i=\Delta_i$: here, PMMMs appear located at the outer QDs \cite{PhysRevB.111.235409}. As for the three‐QD Kitaev chain, we use the extended Nambu basis $\psi = (c_1,\,c_2,\,c_3,\,c_1^\dagger,\,c_2^\dagger,\,c_3^\dagger)$ and find truly localised  Majorana operators ($\bar{\gamma}_{1,3}$) at the outer QDs: $\bar{\gamma}_{1} = (c_{1} + c_{1}^\dagger)/\sqrt{2}$ and $\bar{\gamma_{3}} = i\,(-c_{3}^\dagger + c_{3})/\sqrt{2}$. 
 
That being said, we are interested in exploiting minimal Kitaev chains and their properties for investigating entanglement, which is appropriately described within a many-body formulation. For this reason, in what follows we inspect Eq.\,(\ref{Hamiltonian23a}) and Eq.\,(\ref{Hamiltonian23b}) in a many-body basis.
 
\subsection{Many-body description}
\label{Mbsenergy}
To investigate entanglement and other dynamical properties, Eq.\,(\ref{Hamiltonian23a}) and Eq.\,(\ref{Hamiltonian23b}) are appropriately described in the many-body occupation-number basis. For the two-site chain, we introduce the basis states $\ket{n_1 n_2}$, where $n_i = c_i^\dagger c_i $ represents the number operator that can take values $\{0,1\}$, and $c_i^\dagger$ ($c_i$) creates (annihilates) an electronic state at site $i$. The four basis vectors are $\{\ket{00},\ket{01},\ket{10},\ket{11}\}$, in which the two-site Kitaev chain (2KC) Hamiltonian given by Eq.\,(\ref{Hamiltonian23a})  takes the following form
\begin{equation}
\label{EqenHMat}
\mathcal{H}_{2\mathrm{KC}}=  \left(\begin{array}{cccc}
0 &  0 &  0 & \Delta	\\
0  & \varepsilon_2 & \tau & 0 \\
0 & \tau & \varepsilon_1 &  0 	\\
\Delta & 0  & 0 & \varepsilon_1+\varepsilon_2
\end{array}\right)\,,
\end{equation}
 In this matrix, the odd-parity sector of the Hamiltonian, spanned by $\{\ket{01},\ket{10}\}$, corresponds to the central $2{\times}2$ block, whereas the even-parity sector, spanned by  $\{\ket{00},\ket{11}\}$,  is given by the first and fourth columns/rows. Since the two sectors do not mix, the eigenvalues of the two-site Kitaev chain  in the odd and even parity subspaces can be obtained separately and  given by
\begin{equation}
\label{spectrumEqEnH}
\begin{split}
E_{e}^{\pm}&=\varepsilon_{+}\pm\sqrt{\varepsilon_{+}^{2}+\Delta^{2}}\,,\\
E_{o}^{\pm}&=\varepsilon_{+}\pm\sqrt{\varepsilon_{-}^{2}+\tau^{2}}\,.
\end{split}
\end{equation}
 where $\varepsilon_{\pm} =(\varepsilon_1 \pm \varepsilon_2)/2$.    Moreover, these corresponding eigenstates are given by  
\begin{equation}
\label{eigenstateEqEnH}
\begin{split}
\ket{E_{e}^\pm}
&= N_{e}^\pm 
   \big[-(E_{e}^\pm/\Delta)\,\ket{00}+\ket{11}\big]\,,\\
\ket{E_{o}^\pm} 
&= \frac{1}{\sqrt{2}}\big[\ket{01}\pm\ket{10}\big]\,, 
\end{split}
\end{equation}
where   $N_{e}^\mp =\Delta/\sqrt{\Delta^2 + (E_{e}^{\pm})^{2}}$. 

\begin{figure}
\centering
\includegraphics[width=1\linewidth]{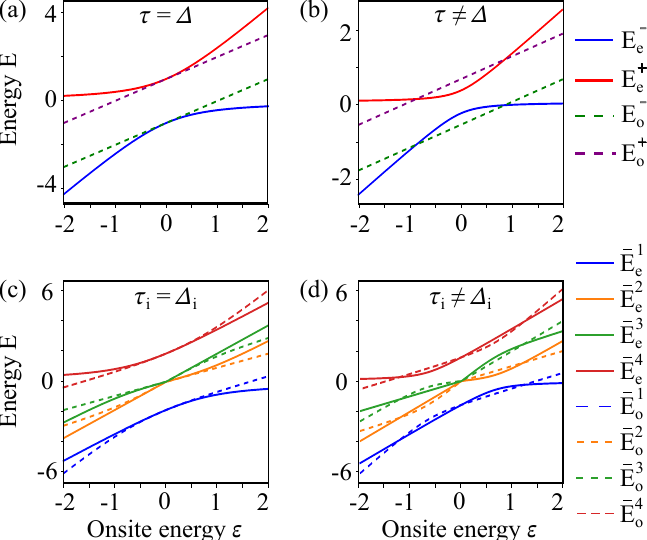}
\caption{Many-body energy spectra of minimal Kitaev chains as a  function of the   onsite energy $\varepsilon_{i}=\varepsilon$ for a two-site  chain (a,b) and a three-site   chain (c,d).  (a) corresponds to the sweet spot ($\Delta=\tau = 1$), while (b) corresponds to the system away from it at $\Delta = 0.5$, $\tau=1$. (c,d) Same as in (a,b), but for a three-site Kitaev chain at the sweet spot ($\Delta_i=\tau_i=1$) and away from it ($\Delta_i=0.5$, $\tau_i=1$).}
\label{energy23sites}
\end{figure}

At  $\varepsilon_{1,2}=0$, the calculation simplifies further, and the eigenvalues and eigenvectors given by Eq.\,(\ref{spectrumEqEnH}) and Eq.\,(\ref{eigenstateEqEnH}) take the form
\begin{equation}
\label{eigenstateweet0EqEnH}
\begin{split}
E_{e}^\pm(\varepsilon_{i}=0)   &= \pm\Delta\,,\\ 
E_{o}^\pm(\varepsilon_{i}=0) &= \pm\tau\,,  
\end{split}
\end{equation}
and 
\begin{equation}
\label{eigenstateweet0EqEnH2}
\begin{split}
\ket{E_{e}^\pm(\varepsilon_{i}=0)} &= \frac{1}{\sqrt{2}}\bigl[\ket{11}\pm\ket{00}\bigr]\,,\\
\ket{E_{o}^\pm(\varepsilon_{i}=0)} &= \frac{1}{\sqrt{2}}\bigl[\ket{10}\pm\ket{01}\bigr]\,,
\end{split}
\end{equation}
Thus, the hopping amplitude $\tau$ controls the energy splitting within the odd-parity sector of the Hamiltonian, while the pairing potential $\Delta$ sets the energies levels  in the even-parity sector. When $\tau = \Delta$, the ground-state energies of both parity sectors coincide $E_{e(o)}^{-}=-\tau$, and such degeneracy signals the emergence of PMMMs as zero energy states ($\delta\equiv |E_{e}^{-}-E_{o}^{-}|=0$) localized one in each QD; this regime is called sweet spot. Moreover, as seen in Eqs.\,(\ref{eigenstateweet0EqEnH2}),  each parity-conserving eigenstate remains maximally entangled. An interesting point to note is that the zero-energy nature of  PMMMs persists even when one of the onsite potentials is detuned from zero \cite{Leijnse2012, PRXQuantum.5.010323}. To visualize the discussion of this part,  the eigenvalues of the even and odd parity sectors given by Eqs.\,(\ref{spectrumEqEnH}) is presented in Fig.~\ref{energy23sites}(a,b). For $\varepsilon_i = 0$ and $\Delta = \tau$, the ground state eigenvalues of both sectors are degenerated but it is lifted when   $\Delta \neq \tau$, as shown in Fig.~\ref{energy23sites}(a,b).

In direct analogy with the two‐site Kitaev chain, we now write the Hamiltonian for a three-site Kitaev chain given by Eq.\,(\ref{Hamiltonian23b}) in the occupation‐number basis $\{\ket{n_1 n_2 n_3}\}$, where $n_i=c_i^\dagger c_i\in\{0,1\}$. Parity conservation again decouples the Hamiltonian into even‐ and odd‐parity sectors, each forming a $4\times4$ blocks. In the even‐parity sector, using the ordered basis $\{\ket{000},\,\ket{011},\,\ket{101},\,\ket{110}\}$, the Hamiltonian is given by
\begin{equation}
\label{Even3HMat}
\mathcal{H}_{3\mathrm{KC}}^{e}=  \left(\begin{array}{cccc}
0 &  \Delta_2 &  0 & \Delta_1 	\\
\Delta_2  & \varepsilon_2+\varepsilon_3 & \tau_1 & 0 \\
0 & \tau_1 & \varepsilon_1 +\varepsilon_3 &  \tau_2 	\\
\Delta_1 & 0  & \tau_2 & \varepsilon_1+\varepsilon_2 
\end{array}\right)\,, 
\end{equation}
which, unlike    the two-site chain, includes both pair potentials ($\Delta_i$) and hopping amplitudes ($\tau_i$)  since all nearest-neighbour processes conserve total fermion parity. We also write down an analogous $4{\times}4$ block for the odd‐parity sector using the ordered basis $\{\ket{001},\,\ket{010},\,\ket{100},\,\ket{111}\}$, given by
\begin{equation}
\label{Odd3HMat}
\mathcal{H}_{3\mathrm{KC}}^{o}=  \left(\begin{array}{cccc}
\varepsilon_3 &  \tau_2 &  0 & \Delta_1 	\\
\tau_2  & \varepsilon_2 & \tau_1 & 0 \\
0 & \tau_1 & \varepsilon_1  &  \Delta_2 	\\
\Delta_1 & 0  & \Delta_2 & \varepsilon_1+ \varepsilon_2+\varepsilon_3
\end{array}\right)\,,
\end{equation}
where, similar to the even sector, it contains both   $\Delta_i$ and   $\tau_i$, which couple nearest-neighbour states while preserving total fermion parity.  

The eigenvalues of $\mathcal{H}_{3\mathrm{KC}}^{e(o)}$, which we denote by $\bar{E}^{i}_{e(o)}$, can be found by solving  ${\rm det}[z-\mathcal{H}_{3\mathrm{KC}}^{e(o)}]=0$ for $z$. In particular,  at $\Delta_{i}=\Delta$, $\tau_{i}=\tau$ they can be found by solving the following polynomial equations
\begin{equation}
\label{EigenvalueEq3SKC}
\begin{split}
z^{4}-6\varepsilon z^{3}+A_{\nu} z^{2}+B_{\nu} z +C_{\nu}&=0\,,\\
%z^{4}-6\varepsilon z^{3}+A_{o} z^{2}+B_{o} z +C_{o}&=0\,,
\end{split}
\end{equation}
where  $\nu=\{e,o\}$ labels the even and odd sector, 
$A_{e}=12\varepsilon^{2}-2\tau^{2}-2\Delta^{2}$, $B_{e}=4\varepsilon \tau^{2}+8\varepsilon\Delta^{2}-8\varepsilon^{3}$, 
$C_{e}=-8\varepsilon^{2}\Delta^{2}$, and $A_{o}=12\varepsilon^{2}-2\tau^{2}-2\Delta^{2}$, 
$B_{o}=8\varepsilon \tau^{2} +4\varepsilon\Delta^{2}-10\varepsilon^{3}$,   $C_{o}=-2\varepsilon^{2}\Delta^{2}-6\varepsilon^{2}\tau^{2}+3\epsilon^{4}$. Once the eigenvalues are found, the eigenstates can be also obtained for each sector.  In the case of the even sector, Eq.\,(\ref{EigenvalueEq3SKC}) can be also written as $(z-2\varepsilon)[z^{3} - 4\varepsilon\,z^{2} 
+ \left(4\varepsilon^{2} - 2t^{2} - 2\Delta^{2}\right)z\,
+\, 4\varepsilon\,\Delta^{2} = 0]$, which gives one eigenvalue given by $\bar{E}^{1}_{e} = 2\varepsilon$, while the remaining $\bar{E}^{2,3,4}_{e}$ are obtained from the cubic equation. In relation to the eigenstates, 
\begin{equation}
\begin{split}
\ket{\bar{E}^{1}_{e}}&=(\ket{000} - \ket{111})/\sqrt{2}\,,\\
\ket{\bar{E}_e^i} &=\; \mathcal{N}(\bar{E}_e^i) \bigg[ 
 \frac{2\Delta}{\bar{E}_e^i} \ket{000} + \ket{011} \\
& + \frac{2\tau}{\bar{E}_e^i - 2\varepsilon} \ket{101} + \ket{110} \bigg],
\end{split}
\end{equation}
where $i=2,3,4$, while $\mathcal{N}(\bar{E}_{e}^i)$ is the normalization constant. 

Simpler expressions are obtained for the eigenvalues when $\varepsilon=0$ and $\Delta=\tau$.  In this case, the even sector eigenvalues are given by 
\begin{equation}
\begin{split}
\bar{E}_{e}^{1(2)}=0\,,\quad \bar{E}_{e}^{3(4)}=\pm\Delta
\end{split}
\end{equation}
with their corresponding normalised eigenvectors given by
\begin{equation}
\label{even3sweetspot}
\begin{split}
\ket{\bar{E}_{e}^1} &= \frac{1}{\sqrt{2}}\bigl[-\ket{000} + \ket{101}\bigr],\\
\ket{\bar{E}_{e}^2} &= \frac{1}{\sqrt{2}}\bigl[-\ket{011} + \ket{110}\bigr],\\
\ket{\bar{E}_{e}^{3,4}} &= \frac{1}{2}\bigl[\mp\ket{000} + \ket{011} \mp \ket{101} + \ket{110}\bigr].
\end{split}
\end{equation}
As seen, these states reflect  that $\ket{\bar{E}_{e}^1}$ is a maximally entangled state of outer QDs in their even fermion parity sector, with the middle QD in the $\ket{0}$ state. In contrast, the state $\ket{\bar{E}_{e}^2}$ is a maximally entangled Bell state of the outer two-QD subsystem in the odd-parity sector, while the middle QD remains occupied, $\ket{1}$, ensuring that the overall three-QD system stays in the even parity sector. The states $\ket{\bar{E}_{e}^{3,4}}$ are linear superpositions of even and odd maximally entangled Bell states of outer QDs, where the middle QD occupies either the $\ket{0}$ or $\ket{1}$ state maintaining the chosen parity of the full Hamiltonian.  Following the same steps, the odd sector energy $\bar{E}_{o}^i$ and its eigenvectors $\ket{\bar{E}_{o}^{i}}$ can be obtained for the three-site Hamiltonian in Eq.~(\ref{Odd3HMat}), see Appendix~\ref{oddDyn3sites} for further details. To further envisage the energies of the even and odd parity sectors in a three-site Kitaev chain, in Fig.\,\ref{energy23sites}, we show them as a function of the on site energies at the sweet spot ($\Delta_{i}=\tau_{i}$) and away from it  ($\Delta_{i}\neq\tau_{i}$); at the sweet spot a degeneracy occurs and PMMMs appear at the outer QDs.

It is worth noting that in the three-site Kitaev chain, at the genuine sweet spot ($\Delta_{i}=\tau_{i}, \varepsilon_{i}=0$), the ground-state energies of the even- and odd-parity sectors of the Hamiltonian are exactly degenerate, see Fig.~\ref{energy23sites}(c). In this \textit{genuine three-site sweet spot}  PMMMs are fully localised in the outer QDs. This degeneracy, and the associated zero-energy PMMMs, can persist even when one or two onsite potentials are tuned to finite values~\cite{Bordin2025}. In this situation, multiple sweet spot configurations can arise depending on the choice of tunneling amplitudes and onsite potentials. For instance, detuning the onsite energy of the central QD ($\varepsilon_2 \neq 0$) reduces the system to an effective two-QD configuration, where the modified regime is referred to as   \textit{effective sweet spot} \cite{Bordin2025,ten_Haaf2025}. Alternatively, when nearest-neighbour  onsite energies are set to finite values ($\varepsilon_2 \neq 0$ and $\varepsilon_3 \neq 0$), the PMMMs become partially delocalised across the chain, in what is referred to as a \textit{delocalised three-site sweet spot}~\cite{PhysRevB.111.235409}. In all cases, however, the degeneracy is lifted when $\Delta \neq \tau$ or when all onsite potentials become nonzero. In this work, we are interested on entanglement properties at and away from the genuine sweet spot, but will discuss alternative sweet spot scenarios whenever relevant.

%%%%%%%%%%%%%%%%%%%%%%%%%%%%%%%%%%%%
%     Section III:  Entanglement dynamics in a two-site Kitaev chain     %
%%%%%%%%%%%%%%%%%%%%%%%%%%%%%%%%%%%%
\section{Entanglement dynamics in a two-site Kitaev chain}
\label{SectionIII}

In this section, we quantify entanglement in a two-site Kitaev chain modelled by Eq.\,(\ref{EqenHMat}), with a particular focus on the sweet spot hosting PMMMs and away from it. For this purpose, we employ several standard measures, such as the concurrence and the geometric measure of entanglement (GME), which are complemented by the return probability and state  entanglement dynamics. While quench-induced entanglement dynamics in superconducting and topological systems \cite{PhysRevA.101.063613, Eisert2015, Heyl_2018, 3wcr-sxtz} have been widely studied—often in the thermodynamic limit-our focus here is on finite, few-site Kitaev chains, where entanglement dynamics are governed by experimentally tunable  parameters  and are directly relevant to QD-Majorana platforms.
 Before going further, we outline how we assess the system dynamics and summarise the definitions and key properties of the entanglement measures employed.

%%%%%%%%%%%%%%%%
%        Subsection III A:           %
%%%%%%%%%%%%%%%%

\subsection{Time-evolved state}
\label{SectionIIIA}
The state dynamics is fundamental to understand how entanglement of the subsystems evolve in time according to a given initial state $\ket{\psi(0)}$. Thus, to address the entanglement dynamics, we need the time-evolved state and is obtained as,
\begin{equation}
\label{tes_num}
    \ket{\psi(t)} = e^{-iHt} \ket{\psi(0)} = \sum_j  e^{-i\lambda_{j} t}\ket{\lambda_j}\bra{\lambda_j}  \ket{\psi(0)}\,,
\end{equation}
where $\ket{\lambda_j}$ and $\lambda_{j}$ are the eigenstates and eigenvalues of the Hamiltonian $H$, respectively. For the two-site Kitaev chain, $H$ can be solved in either parity sector, with the eigenvalues and eigenstates given by Eq.\,(\ref{spectrumEqEnH}) and Eq.\,(\ref{eigenstateEqEnH}), respectively. At the sweet spot $\Delta=\tau$ and $\varepsilon_{i}=0$, the eigenvalues $\lambda_{j}$ and eigenstates $\ket{\lambda_j}$ are given by Eq.\,(\ref{eigenstateweet0EqEnH}) and Eq.\,(\ref{eigenstateweet0EqEnH2}).   

As for the initial states, since a two-site Kitaev chain is fairly small and the fermion parity of the Hamiltonian is conserved, we consider the following representative even-parity initial states, consisting of two separable states followed by two even parity Bell states, respectively,
\begin{equation}
\label{iState}
\begin{split}
\ket{\psi_1(0)} &=  \ket{00}\,, \\
  \ket{\psi_2(0)} &=  \ket{11}\,,\\
 \ket{\psi_3(0)} &=  [\ket{00}+\ket{11}]/\sqrt{2}\,,\\
 \ket{\psi_4(0)} &=  [\ket{00}-\ket{11}]/\sqrt{2}\,.
\end{split}
\end{equation}
As our analysis is mainly restricted to the even-parity sector, only the top two configurations for separable states are allowed. The remaining two states are their maximally entangled superpositions in the same parity sector. We do not list the states with an odd number of quasiparticles, but their form can be easily written down. It is worth noting that, although PMMMs arise from the ground-state degeneracy between the even- and odd-parity sectors of the many-body Hamiltonian, the two-site Kitaev chain Hamiltonian remains block-diagonal in the corresponding parity basis, as evident from the structure of Eq.\,(\ref{EqenHMat}). Thus, the unitary evolution within the even‐parity block alone fully captures the system’s dynamics and entanglement. Under time evolution of the system, we compute the time-dependence of entanglement measures and dynamical observables in the various regimes of the system, which we discuss next.   

%%%%%%%%%%%%%%%%
%        Subsection III B:           %
%%%%%%%%%%%%%%%%

\subsection{Measures of entanglement  in a two-site Kitaev chain}
\label{allquantities}

To investigate and quantify entanglement in a two-site Kitaev chain, we focus on concurrence \cite{ Wootters2001} and the GME \cite{PhysRevA.68.042307}, and complement them by analyzing the return probability and the entanglement dynamics \cite{PhysRevB.110.224510}. Concurrence provides a well-established and experimentally accessible quantifier of bipartite entanglement, suitable for both pure two-site states and reduced mixed states in larger systems \cite{Wootters2001, Wootters1998}. The GME offers a complementary and more general characterization, as it can capture both bipartite and genuine multipartite correlations \cite{PhysRevA.68.042307} and can, in principle, be inferred experimentally \cite{PhysRevLett.92.087902,GUHNE20091}. In the two-site case, the GME is a monotonic function of concurrence \cite{PhysRevA.68.042307}, ensuring a consistent description, while in extended systems it becomes an essential tool for identifying multipartite entanglement \cite{PhysRevA.68.042307}. The return probability and dynamical analysis further elucidate the stability, periodicity, and controllability of the generated entangled states, providing a comprehensive dynamical picture of minimal Kitaev chains.

\subsubsection{Concurrence}
Concurrence is a measure of two-qubit entanglement, applicable to both \textit{pure} and \textit{mixed} states of a given system \cite{Wootters2001}. Since the state of two sites in the two-site Kitaev chain is a \textit{pure state}, it can be written as
 \begin{equation}
\label{2qstate}
\ket{\psi} = \alpha\ket{00}+\beta\ket{01}+\gamma\ket{10}+\delta\ket{11}\,,
\end{equation}
with   $|\alpha|^2+|\beta|^2+|\gamma|^2+|\delta|^2=1$. Then, the concurrence is straight forwardly computed by taking overlap with another state  $\lvert \tilde{\psi}\rangle$ as \cite{Wootters2001}
\begin{equation}
\label{coverlap}
C(\ket{\psi}) = |\langle\psi|\tilde{\psi}\rangle | \,,
\end{equation}
where $\lvert \tilde{\psi}\rangle$  is defined by the spin-flip operation  $\lvert \tilde{\psi}\rangle = (\sigma_y\otimes\sigma_y)\ket{\psi^*}$, in which $\sigma_y$ is the Pauli operator and $\ket{\psi^*}$ is the complex conjugate of the state $\ket{\psi}$. In this situation, the spin-flip operation flips the state $\ket{0}$ to  state $\ket{1}$.  Thus, the spin-flip operation transforms a pure product (separable) state into an orthogonal state, which, by taking the overlap as in Eq.~(\ref{coverlap}), gives   zero. Therefore, the concurrence for a separable state  remains zero, $C=0$ \cite{Wootters2001}. On the other hand, the spin-flip operation on a maximally entangled state does not alter the form of the state, thereby giving a value of $C=1$.  In terms of the coefficients of the state in Eq.\,({\ref{2qstate}), the concurrence can be calculated as 
\begin{equation}
\label{pureC}
C(\ket{\psi}) = 2|\alpha\delta-\beta\gamma | \,.
\end{equation}
Then, the concurrence becomes zero $(C=0)$ when $\alpha\delta = \beta\gamma$. As we will see later, this condition holds particular significance in the two-site Kitaev chain, where it can occur at points of degeneracy. At such points, the energy levels of states are equal, allowing for the same set of coefficients in both the odd-excitation configurations $\lbrace \ket{01}, \ket{10} \rbrace$ and the even-excitation configurations $\lbrace \ket{00}, \ket{11} \rbrace$. The maximum value of concurrence is $C=1$ and can be achieved by setting either $\alpha = \delta = 1/\sqrt{2}$, which then forces $\beta = \gamma = 0$ due to the normalization condition of the state in Eq.~(\ref{2qstate}); alternatively, it can happen for $\beta = \gamma = 1/\sqrt{2}$, which sets $\alpha = \delta = 0$ by the same constraint. In both cases, we obtain the maximally entangled well-known Bell states, either with even or odd excitations, respectively.

\subsubsection{Geometric measure of entanglement}
The concurrence measure can fully characterise a two-qubit system for both separable and entangled initial states \cite{Wootters2001}. However, as the system size increases, the generated entanglement extends beyond bipartite correlations and can no longer be characterised solely by concurrence \cite{Choi2023}. A multipartite entanglement measure becomes necessary, either for larger subsystems or as a global indicator. In this case, the GME is a particularly important tool for quantifying multipartite entanglement, irrespective of the system size \cite{PhysRevA.68.042307}. In what follows, we denote the GME as  $E_G$. The GME is defined in terms of distance: it quantifies how far a given quantum state is from the set of separable states, providing a geometric characterisation of entanglement based on the overlap~\cite{PhysRevA.68.042307}.  Thus, the GME $  E_G$ is obtained as
\begin{equation}
\label{Eg2site}
  E_G(|\psi\rangle) = 1 - \max_{|\phi\rangle} |\langle \phi | \psi \rangle|^2\,,
\end{equation}
where the maximum is taken over all separable states $\ket{\phi}$ in the system’s Hilbert space, and $\ket{\psi}$ represents the state of the system. For the two-site system, as in the two-site Kitaev chain, $E_G$ can be directly expressed in terms of concurrence $C$  as \cite{PhysRevA.68.042307}
\begin{equation}
\label{egc}
  E_G(|\psi\rangle) = \frac{1}{2}(1-\sqrt{1-C^2})\,.
\end{equation}
Hence $E_{G}$ is a monotonically increasing function of $C$: as $C$ varies from $0$ (separable states) to $1$ (maximally entangled states), $E_{G}$ increases from $0$ to $1/2$, vanishing for product (separable) states and reaching $1/2$ for maximally entangled states.  

Before introducing other state-dynamical quantities, we note that related measures, such as quantum discord and quantum mutual information, have been widely used to characterize nonclassical correlations beyond entanglement, particularly in hybrid QD-Majorana systems \cite{Chen_2018, PhysRevB.110.224510, PhysRevB.111.075415}. In this work, however, our focus is on identifying and characterizing highly entangled states in minimal Kitaev chains, for which the measures discussed above, together with the state-dynamical quantities, provide a complete description of the relevant bipartite and genuine multipartite correlations.

\subsubsection{Return probability}
\label{rp}
The dynamics of entanglement can be explored directly via the time-evolved state $\ket{\psi(t)}$ obtained from Eq.\,(\ref{tes_num}). In the two-site Kitaev chain, the Hilbert space splits into even- and odd-parity sectors, and the system oscillates coherently between separable states and maximally entangled Bell states. By projecting the instantaneous state onto either the separable or entangled state, one can directly track its entanglement character \cite{PhysRevB.110.224510}. A particularly useful quantity in this context is the return probability, defined as the overlap between the time-evolved state $\ket{\psi(t)}$ and the initial state $\ket{\psi(0)}$. For an initial pure state $\lvert \psi(0)\rangle$, the return probability is defined as \cite{PhysRevB.110.224510} 
\begin{equation}
\label{RP}
\mathit{R_p} = |\langle \psi(0)|\psi(t) \rangle |^2\,,
\end{equation}
where $\ket{\psi(t)}$ is the time-evolved state of the system. Depending on the chosen initial state, $R_p$ can distinguish whether the system evolves into a separable or an entangled state. Its value ranges from 0 to 1, indicating the degree of separability or entanglement.

\subsubsection{ Entanglement dynamics}
\label{edexpression}
Building on the concept of return probability, we compute the entanglement dynamics to identify instances during the evolution when the system becomes maximally entangled in a particular form. By evaluating this quantity alongside the return probability, we can determine the exact state of the system at the moment of maximum entanglement. This is obtained by projecting the time-evolved state $\ket{\psi(t)}$ onto a maximally entangled target state $\ket{\phi}$, as \cite{PhysRevB.110.224510}
\begin{equation}
\label{ED}
E_d = |\langle \phi|\psi(t) \rangle |^2\,.
\end{equation}
We note that, for the two-site Kitaev chain in the even sector, a maximally entangled target state $\ket{\phi}$ is given by
\begin{equation}
\label{phi_state2}
\ket{\phi} = \frac{1}{\sqrt{2}}[\ket{00}+\ket{11}]\,.
\end{equation}
This choice of $\ket{\phi}$ not unique: one can instead select the orthogonal state obtained by flipping the relative sign of $|11\rangle$ in $|\phi\rangle$, in which case the dynamics remain orthogonal to those of $|\phi\rangle$. Furthermore, computing the return probability for the same initial state $|\phi\rangle$ would yield the same conclusions. Therefore, the combined analysis of these measures fully characterises both entanglement generation and product‐state dynamics.  

In the following part, we present the time-evolution of the entanglement measures discussed in the previous subsections, with a particular focus  on three parameter regimes: i) the sweet spot ($\varepsilon_i=0$ and $\tau=\Delta$); ii)  either $\varepsilon_i=0$ or $\tau=\Delta$; and iii)  a generic case with $\varepsilon_i\neq0$ and $\tau\neq\Delta$. 

%%%%%%%%%%%%%%%%
%        Subsection III C:           %
%%%%%%%%%%%%%%%%

\subsection{Entanglement  at the sweet spot}
Having discussed the measures  to quantify entanglement, in this part we address their dynamics  at the sweet spot ($\Delta = \tau$).  For this purpose, we  make use of the eigenvalues and eigenvectors computed in Eqs.~(\ref{eigenstateEqEnH})  for separable and maximally entangled initial states defined in Eqs.~(\ref{iState}).  We focus on the even sector and, since $\tau$ only appears in the odd sector [Eq.\,(\ref{EqenHMat})], variations in $\tau$ do not affect the dynamics in the even sector.   

\subsubsection{Dynamics of initially separable states}
\label{00dyn}
In the case of a separable initial state in a two-site Kitaev chain without quasiparticles  $\ket{\psi_1(0)}=\ket{00}$,  we obtain the time evolved state given by using Eq.\,(\ref{tes_num}), Eq.\,(\ref{spectrumEqEnH}), and  Eq.\,(\ref{eigenstateEqEnH}). At the sweet spot $\Delta=\tau$ and $\varepsilon_{i}\neq0$, it  reads
\begin{equation}
\label{00dynEqEnH}
\ket{\psi_1(t)}= \alpha_1(t) \ket{00}+ \delta_1(t) \ket{11}\,,
\end{equation}
where $\alpha_1(t)= N_3^2(\lambda_4^2/\Delta^2)e^{-i\lambda_3t} +  N_4^2(\lambda_3^2/\Delta^2)e^{-i\lambda_4t}$, and  $\delta_1(t) = N_3^2(\lambda_4/\Delta)e^{-i\lambda_3t} +  N_4^2(\lambda_3/\Delta)e^{-i\lambda_4t}$. Here, $N_{3}=N_{e}^{-}$ and $N_{4}=N_{e}^{+}$ below Eqs.\,(\ref{eigenstateweet0EqEnH2}). Moreover, $\lambda_{3}=E_{e}^{-}$, and $\lambda_{4}=E_{e}^{+}$, see Eqs.\,(\ref{eigenstateweet0EqEnH}). Then, using Eq.\,(\ref{00dynEqEnH}) and following the discussions in Subsec. \ref{allquantities}, we can calculate the concurrence (C), the geometric measure of entanglement ($E_G$), the return probability ($R_p$), and the entanglement dynamics ($E_d$).  Thus, we obtain
\begin{equation}
\label{SepQuantities}
\begin{split}
C&= 2|\alpha_1(t)\delta_1(t)|\,,\\
E_G& = \frac{1}{2}(1-\sqrt{1-C^2})\,,\\
R_p &= |\alpha_1(t)|^2\,,\\
E_d& = |\alpha_1(t)+\delta_1(t)|^2\,.
\end{split}
\end{equation}
At this point, it is important to make some remarks about Eqs.\,(\ref{SepQuantities}) and Eq.\,(\ref{00dynEqEnH}). The state becomes unentangled whenever either coefficient $\alpha_1(t)$ or $\delta_1(t)$ vanishes, in which case both $C$ and $E_G$ drop to zero. The $R_p$ reaches unity when the system returns to the initial state $\ket{00}$ and vanishes when the state evolves to $\ket{11}$, both of which are separable configurations. In contrast, the vanishing value of $E_d$ depends on the relative phase between $\alpha_1(t)$ and $\delta_1(t)$. Specifically, when $\alpha_1(t) = -\delta_1(t)$, the system evolves into the orthogonal maximally entangled state $(\ket{00} - \ket{11})/\sqrt{2}$, leading to $E_d = 0$. For the maximally entangled state with $\alpha_1(t) = \delta_1(t) = \pm 1/\sqrt{2}$, the observables take the values $C = 1$, $E_G = 0.5$, $R_p = 0.5$, and $E_d = 1$ or $0$, depending on whether the phases of the components are constructive or destructive.

\begin{figure*}[ht]
  \centering
  \includegraphics[scale=0.98]{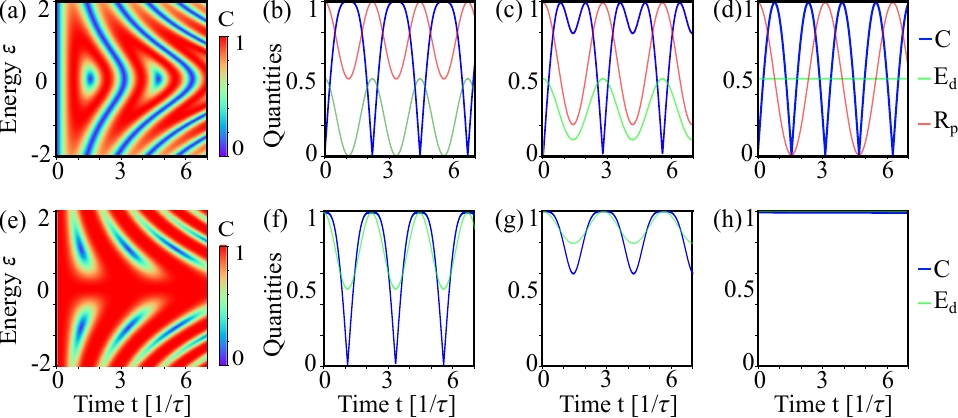}
 \caption{Entanglement measures at the sweet spot ($\tau = \Delta$) for an initially separable state $\ket{00}$ (a-d) and for an initially maximally entangled state $(\ket{00} + \ket{11})/\sqrt{2}$ (e-h). (a) Concurrence ($C$) as  functions of time $t$ and onsite energy $\varepsilon_{i}=\varepsilon$ at the sweet spot ($\tau = \Delta$).  (b)  Time evolution of $C$, return probability $R_p$, and entanglement dynamics $E_d$ at $\varepsilon = 1$. (c,d) Same as in (b), but for  $\varepsilon_1 = 0$, $\varepsilon_2 = 1$ and 
   $\varepsilon_1 = \varepsilon_2 = 0$, respectively. Bottom panel:  (e-h) Same as in (a-d) but for a maximally entangled initial state $(\ket{00} + \ket{11})/\sqrt{2}$.   Parameters: $\Delta=\tau=1$. The geometric entanglement $E_g$ having a monotonic characteristics with $C$ shown in Appendix~\ref{GME2site}. } 
 \label{All00compare}
\end{figure*}

To have a better visualization, in Fig.~\ref{All00compare}(a-d) we present the time-evolution of the entanglement measures given by Eqs.\,(\ref{SepQuantities}) at distinct onsite energies $\varepsilon_{i}$ and at $\Delta=\tau$. First of all, we note that the red regions in $C$ represent maximally entangled states, while the blue regions indicate unentangled states, see Fig.~\ref{All00compare}(a). A similar pattern is exhibited by the geometric entanglement $E_G$, which is a monotonic function of $C$, as given in Eqs.,(\ref{SepQuantities}) (see Appendix~\ref{GME2site} for a complete analysis).
Both quantities, $C$ and $E_G$, exhibit a periodic generation of entanglement, ranging from completely unentangled to maximally entangled states, and show similar qualitative behaviour, as they vary monotonically with respect to each other. Another notable feature of these two measures is that they remain unperturbed when the sign of $\varepsilon$ is reversed. This can be understood from the coefficients in Eq.~(\ref{00dynEqEnH}), accompanied by Eq.~(\ref{spectrumEqEnH}), which remain unchanged under a global sign flip of the onsite energies. During the transition from negative to positive $\varepsilon$, the sweet spot lies at $\varepsilon = 0$, where $C$ exhibits the dynamics of separable and maximally entangled states in a periodic manner over time [Fig.~\ref{All00compare}(a)]. When increasing $\varepsilon$ from 0 to 0.5 nearly doubles the frequency of achieving maximally entangled states [Fig.~\ref{All00compare}(a)], as evident from the number of formed red patches in $C$. To get further insights on the entanglement features at the sweet spot ($\Delta=\tau$), we rewrite all quantities in Eq.~(\ref{SepQuantities}) at $\varepsilon_{i}=0$ by plugging the coefficients defined below Eq.~(\ref{00dynEqEnH}). Hence, 
\begin{equation}
\label{Sweet00Quantities}
\begin{split}
C &= \sin{(2\Delta t)}\,, \\
E_G &= \frac{1}{2}(1-|\cos{(2\Delta t)}|)\,,\\
R_p &= \cos^2{(\Delta t)}\,,\\
E_d & = 1/2\,.
\end{split}
\end{equation}
Here, the periodicity of all measures can be easily identified. Both $C$ and $E_G$ share the same period, given by $t = (2n + 1)\pi / (2\Delta)$, where $n$ is an integer, while $R_p$ exhibits twice this period. Interestingly, $E_d$ remains constant throughout the evolution since in this case it becomes $E_d = |e^{i\Delta t}/\sqrt{2}|^2$, which is why $E_{d}=1/2$ in Eqs.\,(\ref{Sweet00Quantities}) for a separable initial state considered here.

To examine the dynamical characteristics of the entanglement measures in more detail for distinct onsite energies at $\Delta = \tau = 1$, in Fig.~\ref{All00compare}(b-d) we show the time evolution of $C$, $R_p$, and $E_d$ at i) $\varepsilon_{1,2} = 1$; ii) $\varepsilon_{1} =0,  \varepsilon_{2} = 1$; and iii) $\varepsilon_{1,2} = 0$. In the general case with finite onsite energies ($\varepsilon_{1,2}  = 1$),   the concurrence $C$ remains at its maximum value over a finite time interval, producing flat-topped peaks that indicate stable  maximally entangled states, while separable states are reached instantaneously, see blue curve in Fig.~\ref{All00compare}(c).  In this regime, the return probability $R_p$ oscillates between 1 and 0.5, with each minimum coinciding with a maximally entangled configuration ($C = 1$, $E_G = 0.5$). Between these minima, $R_p$ rises to unity, and the system becomes momentarily unentangled, yielding $C =  0$. At these points, the value of $E_d$ reaches 0.5, arising from the contribution of a configuration common to both the initial state and the reference state $\ket{\phi}$ in Eq.~(\ref{phi_state2}).

When only one of the onsite energies is zero, but still away from the genuine sweet spot   ($\varepsilon_{1,2} =  0$, $\Delta = \tau = 1$), one of the PMMMs remains perfectly localised at one QD, while the other PMMM has a finite leakage onto that QD \cite{Leijnse2012}. In this case [Fig.~\ref{All00compare}(c)], the system starts with zero concurrence, which then reaches a maximum, demonstrating the creation of a maximally entangled state. Influenced by the onsite energy, the concurrence forms a valley before reaching another maximum and eventually decays into a separable state that signals the unentangled nature of the system, see the blue curve in Fig.~\ref{All00compare}(c). In relation to the dynamics of $E_d$, it varies arbitrarily for these maximally entangled states, reflecting the relative phase between the two separable configurations $\ket{00}$ and $\ket{11}$, see light green  curve in Fig.~\ref{All00compare}(c). Furthermore,   $R_p$ reaches 1, which indicates a return to the initial separable state where the $C$ vanishes, see orange curve in Fig.~\ref{All00compare}(c).

In the sweet-spot regime ($\varepsilon_{1,2} =  0$, $\Delta = \tau = 1$), Fig.~\ref{All00compare}(d) illustrates a perfectly periodic concurrence $C$ that alternates between zero and its maximum. Unentangled states occur whenever $\cos(\Delta t) = 0$ or $\sin(\Delta t) = 0$ in Eqs.\,(\ref{Sweet00Quantities}), yielding a period of $\pi/2$. Consequently, $C$ reaches its maximum at $t = (2n + 1)\pi/4$ in the genuine sweet spot of a two-site Kitaev chain. Moreover, in this sweet spot regime, the return probability $R_p = \cos^{2}(\Delta t)$ also displays a simple periodic behaviour: it vanishes with period $\pi$, and each interval contains two instants of maximal entanglement. Specifically, the first maximum occurs as $R_p$ evolves from $\ket{11}$, where $R_p = 0$, $C = 0$, $E_G = 0$, to $\ket{00}$ where $R_p = 1$, $C = 0$, $E_G = 0$;  the second maximum appears as the system returns to $\ket{11}$; see orange curve in Fig.~\ref{All00compare}(d) and also Fig.~\ref{All00compare_Eg}(d) in Appendix~\ref{GME2site} for corresponding $E_G$ values. Meanwhile, the entanglement dynamics $E_d$ remains constant since  the relevant coefficients have   real and imaginary parts with equal magnitude of 1/2, see light green curve in Fig.~\ref{All00compare}(d) and Eqs.~(\ref{Sweet00Quantities}). The maximally entangled states generated during the dynamics in the sweet spot regime $\Delta=\tau$ and $\varepsilon_{i}=0$ can be further understood by  writing  the time evolved state using Eq.\,(\ref{00dynEqEnH})  as   $\ket{\psi_1(t)} = (\pm\ket{00} \pm i\ket{11})/\sqrt{2}$, with the relative signs flipping whenever $\sin t$ or $\cos t$ change sign. This unveils that  a two-site Kitaev chain at the sweet spot can generate four types of    maximally entangled states. Nevertheless, irrespective of being or not at the sweet spot, Fig.~\ref{All00compare}(a-d) demonstrate the precise control over entanglement generation, where  maximal entanglement states can be maintained or restore a separable state. This level of control is primarily because of the small system size and the unitary dynamics considered here. In contrast, any coupling to an external environment will generally lead to entanglement decay \cite{PhysRevA.80.062322,delValle:11,PhysRevA.99.042320,Xu_2012}.

Before closing this part, we point out that for the other initial separable state $\ket{\psi_2(0)} = \ket{11}$, the coefficients of the time evolved state at the sweet spot $\Delta=\tau$ and $\varepsilon_{i}\neq0$ are obtained using Eq.\,(\ref{tes_num}), Eq.\,(\ref{spectrumEqEnH}), and  Eq.\,(\ref{eigenstateEqEnH}). Then, $\ket{\psi_{2}(t)}=\alpha_{2}(t)\ket{00}+\delta_{2}(t)\ket{11}$, where  $\alpha_2(t) = -\left[ N_3^2(\lambda_4/\Delta)e^{-i\lambda_3 t} + N_4^2(\lambda_3/\Delta)e^{-i\lambda_4 t} \right]$ and $\delta_2(t) = N_3^2 e^{-i\lambda_3 t} + N_4^2 e^{-i\lambda_4 t}$. Here, $N_{3}=N_{e}^{-}$ and $N_{4}=N_{e}^{+}$ given below Eqs.\,(\ref{eigenstateEqEnH}), while $\lambda_{3}=E_{e}^{-}$, and $\lambda_{4}=E_{e}^{+}$ given by Eqs.\,(\ref{spectrumEqEnH}).  Interestingly, this time evolved state  preserves the orthogonality with respect to the dynamics with initial state $\ket{00}$. Therefore, all entanglement quantifiers exhibit the same features as those for the $\ket{00}$ state discussed above.  These findings therefore highlight the utility  of two-site Kitaev chains for generating highly tunable maximally entangled states by means of  onsite energies.

\begin{figure*}[ht]
  \centering
  \includegraphics[width=0.98\linewidth]{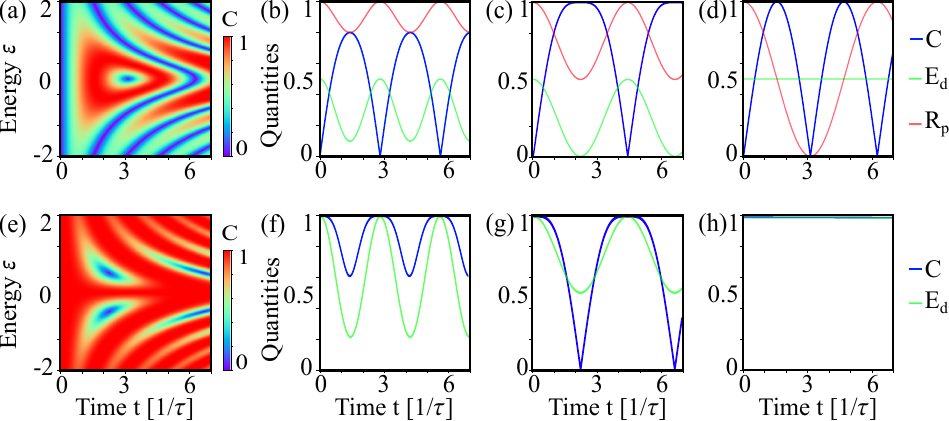}
 \caption{Entanglement measures at  ($\Delta\neq\tau$) for an initially separable state $\ket{00}$ (a-d) and for an initially maximally entangled state $(\ket{00} + \ket{11})/\sqrt{2}$ (e-h). (a) 
  Concurrence ($C$) as functions of time $t$ and onsite energy $\varepsilon_{i}=\varepsilon$ at the sweet spot ($\tau = \Delta$).  (b)  Time evolution of $C$, return probability $R_p$, and entanglement dynamics $E_d$ at $\varepsilon = 1$. (c,d) Same as in (b), but for  $\varepsilon_1 = 0$, $\varepsilon_2 = 1$ and 
   $\varepsilon_1 = \varepsilon_2 = 0$, respectively. Bottom panel:  (e-h) Same as in (a-d) but for a maximally entangled initial state $(\ket{00} + \ket{11})/\sqrt{2}$.   Parameters: $\Delta=0.5$.} 
  \label{Compare00p11eqDT}
 \end{figure*}

\subsubsection{Dynamics of  initially maximally entangled states}
\label{0011dyn}
Following the same steps as in the previous section for a separable initial state, we now consider the dynamics of the maximally entangled initial state $\ket{\psi_{3}(0)}=[\ket{00} + \ket{11}]/\sqrt{2}$ in the sweet spot of a two-site Kitaev chain $\Delta=\tau$, as given in Eq.~(\ref{iState}). The time-evolved state for $\Delta=\tau$ and $\varepsilon_{i}\neq0$ is obtained using Eq.\,(\ref{tes_num}), Eqs.\,(\ref{spectrumEqEnH}), and  Eqs.\,(\ref{eigenstateEqEnH}), which then reads
\begin{equation}
\label{imaxentplus}
\begin{split}
\ket{\psi_3(t)}=\alpha_3(t)\ket{00} + \delta_3(t) \ket{11}\,,
\end{split}                        
\end{equation}
with the coefficients $\alpha_3(t) = N_3^2(\lambda_4/\Delta)e^{-i\lambda_3t}([\lambda_4/\Delta]-1) +  N_4^2(\lambda_3/\Delta)e^{-i\lambda_4t}([\lambda_3/\Delta]-1)$\,, $\delta_3(t) = N_3^2e^{-i\lambda_3t}(1-[\lambda_4/\Delta])  +  N_4^2e^{-i\lambda_4t}(1-[\lambda_3/\Delta])\,$. Here, $N_{3}=N_{e}^{-}$ and $N_{4}=N_{e}^{+}$ and given below Eqs.\,(\ref{eigenstateEqEnH}), while $\lambda_{3}=E_{e}^{-}$, and $\lambda_{4}=E_{e}^{+}$ given by Eqs.\,(\ref{spectrumEqEnH}). The   time-evolved state given by Eq.\,(\ref{imaxentplus}) for the initially maximally entangled configuration $\ket{\psi_{3}(0)}$ takes the same form as that obtained from the separable initial state in Eq.~(\ref{00dynEqEnH}). As a result, the dynamics described by Eq.~(\ref{imaxentplus}) exhibit similar features to those observed for the separable case. However, since the system begins in a maximally entangled state, the time scales and characteristics associated with reaching different separable and entangled states differ. To see these effects, we follow the same steps as for the separable initial state discussed before and obtain the concurrence $C$, entanglement dynamics $E_{d}$, and return probability $R_{p}$. In Figs.~\ref{All00compare}(e), we plot $C$ as a function of time and onsite energy $\varepsilon_{i}\equiv\varepsilon$ at $\Delta=\tau$, while in Figs.~\ref{All00compare}(f-h), we show the time evolution of $C$, $E_{d}$, and $R_{p}$ for i) $\varepsilon_{i}=1$, ii) $\varepsilon_{1}=0$, $\varepsilon_{2}=1$, and iii)  $\varepsilon_{i}=0$. We note that   $E_d$  and  $R_p$ coincide for the chosen initial state, so that only $E_{d}$ is shown. Following the same pattern as in Figs.~\ref{All00compare}(a-d), we do not show the plot of $E_G$ in Figs.~\ref{All00compare}(f-h) for the similar reason.

Since the initial state is maximally entangled, the dynamics begin at unity and evolve periodically. In the density plots of $C$, most regions correspond to strongly entangled states, which is  highlighted by the red areas for $C$ in Figs.~\ref{All00compare}(e). Moreover,  $C$ develops localized  dark-blue patches, which characterize regions of vanishing entanglement. For all finite onsite energies [Fig.~\ref{All00compare}(f)], the system starts in a maximally entangled state with $C = E_d = 1$. Then, the entanglement   periodically drops to zero whenever either $\alpha_3$ or $\delta_3$ vanishes, yielding $C = 0$ and $E_d = 0.5$. This corresponds to the system reaching one of the separable states $\ket{00}$ or $\ket{11}$. Thus, the entanglement between the two QDs in the two-site  Kitaev chain oscillates periodically between separable and maximally entangled states. When only one onsite energy is set to zero [Fig.~\ref{All00compare}(g)], the system no longer reaches fully separable states. Instead, it oscillates between a maximally entangled state and partially entangled configurations. This results in nonzero minima for $C$, consistent with the behavior of $E_d$, which does not reach 0.5. Finally, when both onsite energies are set to zero [Fig.~\ref{All00compare}(h)], the initial state becomes an eigenstate of the Hamiltonian. Consequently, the system evolves only by a global phase, and the time-evolution is simply   constant  in all the entanglement measures, see Fig.~\ref{All00compare}(h).

\subsection{Entanglement away from the sweet spot}
\label{subsectionIIID}
For $\Delta \neq \tau$, the two-site Kitaev chain does not host PMMMs but the system still possesses  interesting entanglement properties due to the $p$-wave symmetry of the pair potential. We note that   $\tau$ does not affect  the even sector that we are exploring and the analysis and expressions are exactly the same as in the previous subsection. The only difference with the previous subsection is that here the entanglement measures  for initially separable and maximally entangled states are obtained at  a smaller value of $\Delta$. 

\subsubsection{Dynamics of initially separable states}
We start by considering the initial separable state $\ket{00}$ and obtain all the entanglement measures discussed in the previous subsection. In Figs.~\ref{Compare00p11eqDT}(a), we present $C$ as a function of time and onsite energy $\varepsilon_{i}=\varepsilon$ at $\Delta=0.5$. Moreover, in Figs.~\ref{Compare00p11eqDT}(b-d), we present the time evolution of $C$, $E_d$, and $R_p$ for i) $\varepsilon_{i}=1$; ii) $\varepsilon_{1}=0$, $\varepsilon_{2}=1$; and iii) $\varepsilon_{i}=0$. As expected, the dynamics begins from a state with zero entanglement and evolves periodically [Figs.~\ref{Compare00p11eqDT}(a-d)], following a similar behavior as for $\Delta=1$ in Fig.\,\ref{All00compare}(a-d) but with slightly  modified features. In particular, $C$ develops broad regions with  intermediate levels of entanglement but also regions with high entanglement, see red regions in  Figs.~\ref{Compare00p11eqDT}(a). The areas with high entanglement emerge around zero onsite energies $\varepsilon=0$,   with a periodic time evolution between high and vanishing entanglement at $\varepsilon=0$ [Figs.~\ref{Compare00p11eqDT}(a)]. By fixing the onsite energies at $\varepsilon_{i}=1$, the time evolution of  $C$ for the separable initial state  exhibits pronounced periodic peaks but never reaches unity throughout the evolution [Figs.~\ref{Compare00p11eqDT}(b)]. This indicates the formation of highly entangled states without achieving a fully maximally entangled state.  This behaviour is further supported by $E_d$, which starts from 0.5 since one configuration of the entangled basis state $\ket{\phi}$ overlaps with the initial state, and it never drops to zero because  the amplitudes of $\ket{00}$ and $\ket{11}$ in Eqs.~(\ref{SepQuantities}) never become equal in magnitude and opposite in sign at any point. The return probability $R_p$ in this case also oscillates near unity, indicating that the achieved state approaches the initial state $\ket{00}$; here, for a maximally entangled state, the configuration $\ket{11}$ would be needed to realize, but it is not being generated during the time evolution in Figs.~\ref{Compare00p11eqDT}(b).  Therefore, in contrast to the sweet spot regime discussed in the previous subsection, the  regime with smaller $\Delta$ and $\varepsilon_{i}\neq0$ cannot generate maximally entangled states in two-site Kitaev chains.

Interestingly, tuning one of the onsite energies to zero enables the system to generate maximally entangled states, as presented in Fig.~\ref{Compare00p11eqDT}(c). In contrast to the   case with finite onsite energies, here $R_p$ drops from 1 to 0.5 and $C$ attains unity when a maximally entangled state is formed. Tuning both onsite energies to zero evolves the system similarly to the sweet spot case, where it oscillates periodically between separable and maximally entangled states, with the period determined by $\Delta$, see Fig.~\ref{Compare00p11eqDT}(d). Thus, in two-site Kitaev chains with even smaller values of $\Delta$, which can put the system away from the sweet spot, it is possible to generate maximally entangled states from initially separable states by simply tuning the onsite energies.

\subsubsection{Dynamics of initially maximally entangled states}
In the case of the initial maximally entangled state $(\ket{00} + \ket{11})/\sqrt{2}$, Figs.~\ref{Compare00p11eqDT}(e) shows $C$ as a function of time and onsite energy $\varepsilon_{i}=\varepsilon$ at $\Delta=0.5$. Also, in Figs.~\ref{Compare00p11eqDT}(f-h) we present the time evolution of $C$,
$E_d$, and $R_p$ for distinct onsite energies $\varepsilon_{i}$; note that  $R_p$ is not shown because it is identical to $E_d$ in this case. Since the initial state is maximally entangled, the dynamics begins at unity and evolves periodically, with a period that is longer than the case with $\Delta=1$. We further notice that  $C$ exhibits very large regions with highly entangled states, with entanglement vanishing only over a narrow range of onsite energies, see red  and dark blue patches in Figs.~\ref{Compare00p11eqDT}(e). When looking at the time evolution of the entanglement measures at $\varepsilon_{i}=1$ [Figs.~\ref{Compare00p11eqDT}(f)], the first feature we see is that  $C$ and $E_G$  never drop to zero, unlike the case with $\Delta=1$ in 
Fig.~\ref{All00compare}(f). Instead, they exhibit arbitrary minima, a behavior that is similar to Fig.~\ref{All00compare}(g) but in a distinct parameter regime.  Interestingly, $E_d$ drops to a value below 0.5, indicating a destructive phase shift between the two components of the dynamical state [Fig.~\ref{All00compare}(g)]. It is also noteworthy that $E_d$ never falls to zero before returning to 1, implying that no other maximally entangled state is formed during the evolution.

Tuning one of the onsite energies to zero enables the system to reach separable states, as illustrated in Fig.~\ref{Compare00p11eqDT}(g). In contrast to the previous case, here $E_d$ drops from 1 to 0.5, and $E_G$ reaches zero, with $C$ also indicating that the system transitions into a separable state. However, $E_d$ does not reach zero, suggesting that only the maximally entangled state of the form $(\ket{00} + \ket{11})/\sqrt{2}$ is dynamically accessible. When both onsite energies are set to zero [Fig.~\ref{Compare00p11eqDT}(h)], the behaviour resembles the sweet spot dynamics for an initially maximally entangled state [Fig.~\ref{All00compare}(h)]. The initial state becomes an eigenstate of the Hamiltonian, leading to  time evolution with a global phase $\ket{\psi_3(t)} = e^{-i\lambda_3 t}\ket{\psi_3(0)} = e^{-i\Delta t}[\ket{00} + \ket{11}]/\sqrt{2}$ generating a constant time evolution. At this point, we would like to remark that all the results presented in this subsection correspond to $\Delta=0.5$, implying $\Delta<\tau$. When $\Delta>\tau$, the periodicity in the dynamics of the entanglement measures exhibit denser regions with highly entangled states, and even maximally entangled states. 

In summary, we conclude that minimal two-site Kitaev chains can realize maximally entangled states with a considerable degree of tunability by the onsite energies and pair potential. Since large values of $\Delta$ are very likely to promote broader regions with stable maximally entangled states,  it will be necessary to enhance and control CAR processes \cite{bordin2023tunable,bordin2024crossed}.

\section{Entanglement dynamics in a three-site Kitaev chain}
\label{SectionIV}
As we have seen in the previous section, the dynamics in a two-site Kitaev chain typically oscillates between separable and maximally entangled states when starting from an unentangled configuration. Initialising the system in a maximally entangled state leads to a different behaviour, but, in both cases, the evolution consistently features highly entangled states. This property is an outcome of the system’s minimal size and direct coupling between  QDs. It is then natural to wonder whether  entanglement generation persists in a three-site system, where PMMMs reside on the outer QDs and are mediated by a central QD. 

As discussed in Section \ref{SectionII}, several parameter regimes support PMMMs in a three-site Kitaev chain, which leads to distinct types of sweet spots. We remind that the genuine sweet spot is achieved when the pair potentials $\Delta$ is equal to the hopping amplitudes and the onsite energies are set to zero: $\Delta_{i}=\tau_{i} $ and $\varepsilon_{i}=0$ \cite{Bordin2025}. Moreover, an effective sweet spot occurs when, additionally to the condition for the pair potentials and hopping amplitudes being equal, the middle QD energy is away from zero: $\Delta_{i}=\tau_{i}$, $\varepsilon_{2}\neq0$, $\varepsilon_{1,3}=0$ \cite{Bordin2025,ten_Haaf2025}. Yet another possibility is to have a delocalised three-site sweet spot which occurs when only the outer onsite energies are non zero and the pair potentials are equal to hopping amplitudes:   $\Delta_{i}=\tau_{i}$, $\varepsilon_{2}=0$, $\varepsilon_{1,3}\neq0$ \cite{PhysRevB.111.235409}. Thus, several parameter regimes support PMMMs in a three-site Kitaev chain. Since we are interested in the generation and characterization of entanglement in this system, we first follow the procedure outlined in Subsec.~\ref{SectionIIIA} to obtain the time-evolved state of the three-site Hamiltonian. We then compute the entanglement measures introduced in Subsec.~\ref{allquantities} across various parameter regimes. Since these quantities take on different interpretations in the context of the three-site Kitaev chain, we start by briefly revisiting  their definitions.

\subsection{Measures of entanglement in a three-site Kitaev chain}
\label{Quantities3sites}

When the system size increases beyond the two-site case, the reduced two-site state generally becomes mixed, which makes the evaluation of entanglement measures in a three-site Kitaev chain more challenging than in the two-site case. This, in turn, implies that both the qualitative and quantitative interpretation of entanglement measures can change. To assess entanglement in this section, we therefore employ the entanglement and state-dynamical measures introduced in Sec. III. We begin by analyzing the concurrence, which remains a suitable bipartite measure when applied to reduced subsystems of larger systems.

\subsubsection{Concurrence}
\label{concurrence3SKC}
In quantum systems with more than two sites, the state of a smaller part—such as a pair of sites—is typically not pure but mixed, as it includes the influence of the rest of the system \cite{nielsen2010quantum, PhysRevLett.130.190401, v7gb-5gq8}.   In this case, $C$ characterises entanglement between two qubits in a mixed state \cite{Wootters2001}. Thus, the concurrence $C$ between a pair of sites in the three-site Kitaev chain can be computed using the two-site reduced density matrix (RDM)  \cite{RevModPhys.81.865, PhysRevA.98.052303, PhysRevA.102.012406}. The RDM ($\rho_d$ ) is obtained by tracing out the remaining degrees of freedom from the composite state of the system as \cite{nielsen2010quantum}
\begin{equation}
\label{reducedDM}
    \rho_{d} = Tr_{{d}'}(\rho)\,,
\end{equation}
where $Tr_{d'}$ denotes the trace over the sites other than $d$ part of the system and $\rho = \ket{\Psi}\bra{\Psi}$ is the density matrix for state  $\ket{\Psi}$   of the system. Then, for more than two-site systems, the concurrence is given in terms of its eigenvalues of the RDM as \cite{Wootters1998}
\begin{equation}
\label{Concur}
C_{ij} \,=\, \max\{\eta_1 - \eta_2 - \eta_3 - \eta_4,0\}\,,
\end{equation}
where the right-hand expression inside the bracket corresponds to the decreasing order of the eigenvalues $\eta_i$ of the matrix $Q = \sqrt{\sqrt{\rho_{d}}\;\tilde{\rho}_{d}\;\sqrt{\rho_{d}} }$. The matrix $\tilde{\rho_d}$ is defined as $\tilde{\rho_d}= \sigma_y\, \otimes\, \sigma_y\, \rho_d^{*}\, \sigma_y\,\otimes\,\sigma_y$, where $\rho_d^{*}$ is the complex conjugate of $\rho_d$ given by Eq.~(\ref{reducedDM}).  Then, the RDM in the two-site basis  $\{\ket{00}$, $\ket{01}$, $\ket{10}$, $\ket{11}\}$ reads
\begin{equation}
\label{rdm1}
\rho_{d}= 
\begin{pmatrix}
\rho_{11} & 0 & 0 & \rho_{14}	\\
0  & \rho_{22} & \rho_{23} & 0 \\
0  & \rho_{32} & \rho_{33} &  0	\\
\rho_{41} & 0  & 0 & \rho_{44} \\
\end{pmatrix}\,,
\end{equation}
where each nonzero element of the RDM can be expressed in terms of the coefficients of the system’s state $\ket{\Psi}$. The vanishing elements reflect fermion parity conservation of the Hamiltonian, indicating that the system resides in either the even or odd sector without mixing them. The diagonal terms capture classical correlations, while the counter-diagonal elements encode quantum coherence. Also, the RDM has an $X$-form \cite{ PhysRevA.98.052303} and, in this case, the concurrence can be computed analytically in terms of the matrix elements of the RDM as \cite{Hedemann2018}
\begin{equation}
\label{concFormula}
    C_{ij} = 2{\rm max} \lbrace 0, |\rho_{23}|- \sqrt{\rho_{11}\rho_{44}}, |\rho_{14}|- \sqrt{\rho_{22}\rho_{33}} \rbrace\,,
\end{equation}
where the maximum of the three elements is found by obtaining the eigenvalues of the matrix $Q$ below Eq.\,(\ref{Concur}). It is worth noting that, in Eq.\,(\ref{concFormula}), a nonzero concurrence requires dominant off‐diagonal  components in the RDM;  these components arise from the coherent excitation of specific basis configurations and are essential for generating entanglement. 

We also point out that, in the three-site Kitaev chain that we focus on in this part, there are three distinct pairs of QDs for which concurrence can be evaluated. The concurrences $C_{12}$ and $C_{23}$, corresponding to the first and second QDs and the second and third QDs, respectively, quantify what we refer to as nearest-neighbour entanglement. In contrast, the concurrence $C_{13}$, between the first and third QDs, captures next-nearest-neighbour entanglement. All of these concurrences between two QDs will be able to identify bipartite entanglement. As we will show later, these different concurrences reveal important features of three-site Kitaev chains  and highlight both similarities and differences in entanglement dynamics compared to the two-site system studied in the previous section.

\subsubsection{Geometric measure of entanglement}
\label{gm3sites}
As the system size increases beyond two sites, the generated entanglement is no longer purely bipartite, and concurrence alone becomes insufficient to fully characterise the entanglement present. In particular, multipartite entanglement naturally arises, especially in our case, where PMMMs may overlap either with one another or with the central QD. To capture the features of multipartite entanglement in the three-site Kitaev system, we employ the symmetric-ansatz estimate of GME ($E_{\rm G}$)~\cite{PhysRevA.68.042307}. Unlike in the two-site case, where the GME coincides with concurrence and captures bipartite correlations, here we will explicitly employ it  to quantify multipartite entanglement. The basic definition follows that written for the two-site Kitaev chain    in Eq.~(\ref{Eg2site}), which we rewrite for completeness $ E_G(|\psi\rangle) = 1 - \max_{|\phi\rangle} |\langle \phi | \psi \rangle|^2$; here  $\ket{\phi}$ denotes a separable state in the Hilbert space of the three-site system, and $\ket{\psi}$ is the system’s state. For our calculation, The set of separable states in a three-site system is defined as in Ref.~\cite{PhysRevA.68.042307},
\begin{equation}
\label{sep3state}
 \ket{\phi} = (\cos\theta\ket{0}+\sin\theta\ket{1})^{\otimes 3}\,,
\end{equation}
 which is symmetric product states of three qubits up to local phase transformations of the form $\lbrace \ket{0}, \ket{1} \rbrace \rightarrow \lbrace \ket{0}, e^{i\phi}\ket{1} \rbrace$.  With this setup, we use Eq.\,(\ref{Eg2site}) and calculate  $E_{\rm G}$ for two characteristic entangled states of three qubits. For permutation-symmetric states such as the GHZ and W states, the symmetric-product-state ansatz serves a natural and well-motivated choice for estimating the multipartite entanglement\cite{PhysRevA.68.042307}. In particular, we consider the GHZ state $
\ket{\text{GHZ}} = \left( \ket{000} + \ket{111} \right)/\sqrt{2}$, which was shown to be crucial in metrology \cite{giovannetti2011advances,PhysRevLett.134.130604}, and also the W state in the even sector $\ket{{\rm W}_{e}} =  \left( \ket{011} + \ket{101} + \ket{110} \right) /\sqrt{3}$. Before going further, we highlight that the GHZ state is a genuinely multipartite entangled state, while the W state can exhibit both bipartite and multipartite entanglements.  Furthermore, the W state is symmetric under permutation and robust to qubit loss for its excitations either in odd or even sectors, or the superposition of both. This makes the W state  robust for the entanglement under the loss of one qubit, which means the two remaining qubits remain entangled even when one qubit is lost.  Specifically, in practical three-qubit devices, it is challenging to realize the perfect W state   $\ket{{\rm W}_e}$. Instead, due to parity conservation of the Hamiltonian, the system is expected to generate   superpositions of $\ket{000}$ and $\ket{W_e}$, which leads to an imperfect W state. In this case, even an imperfect W state   $\ket{\bar{W}} = \alpha \ket{000} + \beta \ket{W_e}$ can retain sufficient multipartite entanglement to outperform any purely bipartite or separable strategy in quantum protocols; this is because $\ket{000}$ can maintain  multipartite correlations. Due to these reasons, we here inspect GME for the GHZ and imperfect  W state. 
 
With the separable state $\ket{\phi}$ given by  Eq.\,(\ref{sep3state}), we now calculate   $E_{\rm G}$ for the  GHZ and imperfect W states ($\ket{\text{GHZ}}$ and $\ket{{\rm W}_{e}}$)   following  Eq.\,(\ref{Eg2site}). This requires to   maximise  the magnitude of overlaps   $\bra{{\rm GHZ}}\ket{\phi}$ and $\bra{{\rm W}_{e}}\ket{\phi}$ which are given by
\begin{equation}
\label{overlapf}
\begin{split}
  |\bra{{\rm GHZ}}\ket{\phi}| &\equiv   \Lambda^{\rm GHZ} = |\frac{1}{\sqrt{2}}[\cos^3(\theta) + \sin^3(\theta)]|\,,\\
    |\bra{{\rm W}_{e}}\ket{\phi}| &\equiv   \Lambda^{\rm W} =|\frac{1}{\sqrt{3}}[\cos^2(\theta)\sin(\theta)]|\,,
  \end{split}
\end{equation}
over $\theta\in[0, \pi]$.  In doing so, we find  $\Lambda_{max}^{\rm GHZ}={\rm max}_{\theta}
\Lambda^{\rm GHZ} = 1/\sqrt{2}$  that gives the value of the GME for the GHZ state as $E^{\rm GHZ}_{\rm G}=1/2$. For  the perfect W state, we obtain  $\Lambda_{max}^{W} ={\rm max}_{\theta}
\Lambda^{\rm W} = 2/3$ which gives $E^{\rm W}_{\rm G}=5/9$,  which exceeds the two-qubit upper bound $E_G \leq 1/2$ [see discussion below Eq.~(\ref{egc})].   This also infers that in a three-qubit system, as it is the three-site Kitaev chain considered in this part, the value of the GME  for the W state can go up to $E^{\rm W}_{\rm G}=1$. These features correspond of GME for W and GHZ states under ideal three-qubit conditions. However, in more realistic systems such as the three-site Kitaev chain considered here, it is not always possible to dynamically achieve a truly W state~\footnote{A pure W state can be created in three-qubit systems under certain circumstances  \cite{PhysRevA.62.062314, PhysRevA.68.042307},  where   neither $\ket{000}$ nor $\ket{111}$ appears in the eigenstates, and where the initial state also lacks these configurations. The system must then evolve under dynamics that preserve these conditions.}. This occurs because, during time evolution, all accessible configurations within a given parity sector of the Hamiltonian contribute to the dynamics. As a result, the time-evolved state generally becomes a superposition of all configurations within one or both parity sectors of the Hamiltonian, depending on the choice of initial state. Therefore, the resulting time-evolved state is more accurately described by the general expression that combines both parity sectors of the Hamiltonian, see Eq.~(\ref{t_aribrary3qubitstate}).

Before ending this part,  we remind that concurrence alone may vanish for certain pairs, falsely suggesting the absence of entanglement, whereas the GME remains positive as long as $\beta \neq 0$. This ensures, for example, a quantum advantage in controlled teleportation~\cite{YeoChua2006} or quantum secret sharing~\cite{Hillery1999}, and provides enhanced sensitivity in distributed metrology~\cite{RevModPhys.90.035005}. Moreover, by tracking how the GME varies with the admixture parameter \(\alpha\), one can optimise the trade‐off between robustness to qubit loss (favouring more $\ket{000}$) and entanglement‐powered performance, making GME a directly applicable figure of merit when engineering three‐qubit entangled resources in noisy, near‐term hardware. For these reasons, we explore the GME in a three-site Kitaev chain.

\subsubsection{Return probability and entanglement dynamics}
\label{RpEd3SKC}
To find the return probability $R_{p}$ and entanglement dynamics $E_{d}$, we employ the same expressions as those used in Subsec.~\ref{rp} and given by Eq.\,(\ref{RP}) and Eq.\,(\ref{ED}), respectively. While the calculation is the same as before, we remark some important points that need to be taken into account during the calculation. First, the definition of   $R_{p}$ depends only on the overlap between the time-evolved and initial states but it is independent of the system size; thus, we must be careful when defining the initial state, see the next Subsection. Second, the calculation of $E_{d}$ depends on the choice of the target state $\ket{\phi}$, which must be redefined for the three-site system.   Following the analogy with the two-site Kitaev system, we consider an entangled target state of the form
\begin{equation}
\label{phi_state3}
\begin{split}
\ket{\phi_{1}} &= \frac{1}{\sqrt{2}}\left(\ket{000} + \ket{111}\right)\,,\\
\ket{\phi_{2}} &= \frac{1}{\sqrt{2}}\left(\ket{000} + \ket{101}\right).
\end{split}
\end{equation}
The first target state represents a superposition between the empty   and the  fully occupied states, while the second state represent a maximally entangled state between first and third QDs when the middle QD is in the $\ket{0}$. Note that $\ket{\phi_{1}}$ may appear to be a natural extension of the two-site case, it differs qualitatively: the configurations $\ket{000}$ and $\ket{111}$ belong to different parity sectors of the Hamiltonian: $\ket{000}$ lies in the even sector, while $\ket{111}$ lies in the odd sector. As a result, computing $E_d$ in the three-site Kitaev chain   requires contributions from both even and odd parity sectors of the Hamiltonian. Another feature of the entangled state in Eq.~(\ref{phi_state3}) is that  it is the GHZ state, which we discussed in the previous subsection. Therefore, $E_d$ in the three-site chain quantifies the probability that the system evolves into the GHZ configuration during its dynamics.

\subsubsection{Initial states in a three-site Kitaev chain}
To investigate both bipartite and multipartite entanglement in the three-site Kitaev chain, we consider two distinct initial states
\begin{equation}
\label{istate3}
\begin{split}
 \ket{\psi_5(0)} &= \ket{000}\,,\\
 \ket{\psi_6(0)} &= \frac{1}{\sqrt{2}} \left( \ket{000} + \ket{111} \right)\,.
\end{split}
\end{equation}
The first state, $\ket{\psi_5(0)}$, is a fully separable product state and serves as a natural extension of the two-site system. It is well-suited for studying the emergence of both bipartite and multipartite entanglement under time evolution. The second state, $\ket{\psi_6(0)}$, is the GHZ state discussed in the previous subsection and is a canonical example of genuine multipartite entanglement in three-qubit systems.This choice enables us to probe the geometric and nonlocal aspects of entanglement that extend beyond bipartite correlations. In the three-site Kitaev chain, analyzing entanglement for separable and initially maximally entangled states is not as straightforward as in the two-site case discussed in Sec.~\ref{allquantities}. Therefore, in what follows we present the results at and away from the sweet spot together to facilitate a direct comparison.

\subsection{Entanglement at and away from the sweet spot}
As discussed earlier, the three-site Kitaev chain contains multiple  pairs of sites that allow us to quantify both local and nonlocal concurrence. In this section, we present a step-by-step calculation of each concurrence at the genuine sweet spot, where all pairs individually satisfy the sweet spot condition. Alongside this analysis, we also comment on scenarios where one or more onsite energies are finite, leading to delocalisation of PMMM wave functions. At the genuine sweet spot, defined by $\varepsilon_i = 0$ and $\tau_i = \Delta_i$, we use the eigenvalues and eigenstates of the Hamiltonian given in Eq.~(\ref{even3sweetspot}) to construct the time evolution of an initial state as described in Sec.~\ref{SectionIIIA}.

\begin{figure*}
  \centering
     \includegraphics[width=0.95\linewidth, height=0.7\textheight, keepaspectratio]{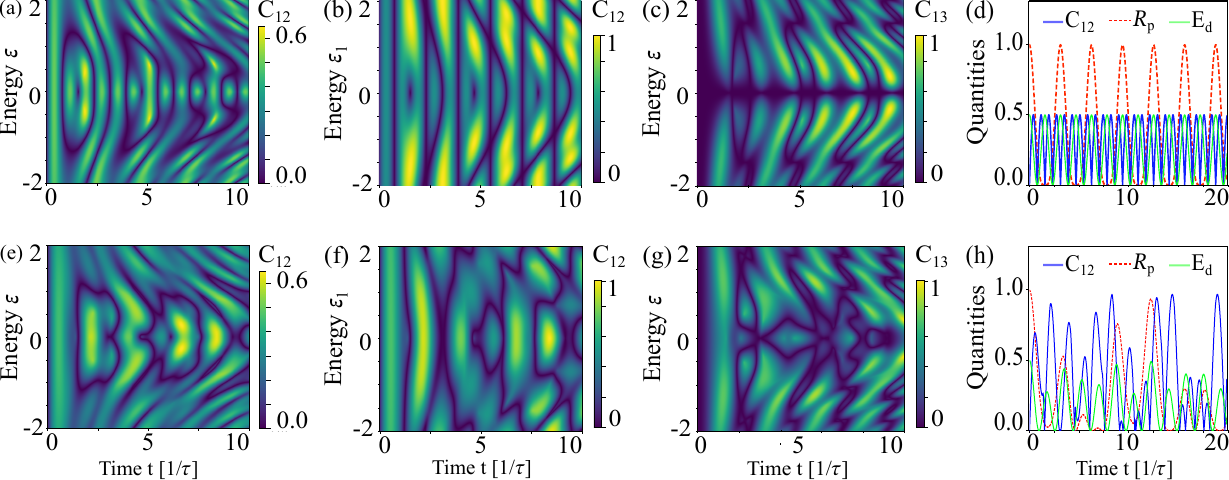}
 \caption{(a) Concurrence between the first and second QDs ($C_{12}$) as a function of time and onsite energies $\varepsilon\equiv\varepsilon_i$ for an initial separable state $\ket{000}$ in a three-site Kitaev chain. (b) Same as in (a), but as a function of time and  $\varepsilon_{1}$ at $\varepsilon_{2,3}=0$. (c) Concurrence between the first and third QDs ($C_{13}$) as a function of time and onsite energies $\varepsilon\equiv \varepsilon$ for the same conditions of (a). (d) Time evolution of $C_{12}$, $R_{p}$, and $E_{d}$ at the genuine sweet spot $\Delta_{i} = \tau_{i}$ and $\varepsilon_{i} = 0$ for the same initial state  of (a). (e-h) Same quantities as in (a-d), but at $\tau_{1,2} \equiv  \Delta_{1} = 1$ and $\Delta_2 = 0.5$, for the initial state of  (a).}
\label{C12_13commonEDensity}
\end{figure*}

\subsubsection{Initial separable state}
\label{evenDynsep3KC}
In this section, we compute the dynamics for the initially separable state  $\ket{\psi_5(0)} = \ket{000}$ given in Eq.~(\ref{istate3}). The time evolution of the system can be expressed using the spectral decomposition of the Hamiltonian as $\ket{\psi(t)} = \sum_{i=1}^3 e^{-iE_e^{i} t} \ket{E_e^{i}} \bra{E_e^{i}}\ket{ \psi(0)}$,
as derived from the eigenstates and eigenvalues in Eq.~(\ref{even3sweetspot}). Only three eigenstates contribute to the dynamics because the fourth eigenstate is orthogonal to the initial state, i.e., $\bra{E_e^{4}}\ket{\psi(0)} = 0$, and thus does not participate in the time evolution. Further simplification using the explicit forms of the eigenvalues and eigenvectors yields the following expression
\begin{equation}
\label{psit000}
\begin{split}
    \ket{\psi(t)} &= \cos^2(\Delta t)\ket{000} - \sin^2(\Delta t)\ket{101} \\
                        & - \frac{i}{2}\sin(2\Delta t)[\ket{011}+ \ket{110}] \,.
\end{split}
\end{equation}
From the above time-evolved state, we compute the relevant two-site concurrences for the three-site Kitaev chain following the discussion of Subsec.\,\ref{concurrence3SKC}. These include nearest-neighbour and next-nearest-neighbour concurrences between nearest-neighbour pairs (sites 1–2), as well as next-nearest-neighbour concurrence between the outer QDs (sites 1 and 3). Notably, the time-evolved state given by Eq.\,(\ref{psit000}) lies in the even parity sector of the Hamiltonian; an analogous analysis can be performed for the odd parity sector. Since the Hamiltonian conserves parity, the dynamics within each parity sector evolve independently, allowing a complete description to be obtained by analysing either sector separately. We will return to the role of both sectors when introducing the GME for characterising multipartite correlations. For now, we focus on bipartite entanglement by evaluating the nearest-neighbour and next-nearest-neighbour  concurrences in the dynamical state given in Eq.~(\ref{psit000}).

\subsubsection{Nearest-neighbour and next-nearest-neighbour  concurrences}
In this section, we quantify  two-site concurrences between different pairs representing   entanglement between nearest-neighbour and next-nearest-neighbour  QDs. For this purpose, and  following Subsec.\,\ref{concurrence3SKC}, we employ the nearest-neighbour reduced density matrix (RDM) $\rho_{12}$ obtained from from the time-evolved state given in Eq.~(\ref{psit000}). Thus, we obtain
\begin{equation}
\label{rdmpsit12}
\rho_{12}= 
\begin{pmatrix}
|\alpha_1|^2 & 0 & 0 & \alpha_1\delta_1^*	\\
0  & |\beta_1|^2 & \beta_1\gamma_1^* & 0 \\
0  & \beta_1^*\gamma_1 & |\gamma_1|^2 &  0	\\
\alpha_1^*\delta_1 & 0  & 0 & |\delta_1|^2 \\
\end{pmatrix}\,,
\end{equation}
where the matrix elements are given by the coefficients of Eq.~(\ref{psit000}), with $\alpha_1 = \cos^2(\Delta t)$, $ \gamma_1 = -\sin^2(\Delta t)$, and $\beta_1 = \delta_1 =i\sin(2\Delta t)/2$. Then, using these coefficients in $\rho_{12}$ and the concurrence formulae in Eq.~(\ref{concFormula}), we can write concurrence as
\begin{equation}
\label{C12}
    C_{12} = \max \lbrace 0,  |\sin(4\Delta t)|/2\rbrace\,,
\end{equation}
where   the pre-factor   $1/2$ in the second element restricts the maximum concurrence to be $1/2$. Moreover, the time period of the concurrence oscillation is given by $T = \pi / 4\Delta$. At this time, the time-evolved state of the system [Eq.\,(\ref{psit000})] takes the form $[\ket{000} + i\ket{011} + \ket{101} + i\ket{110}]/2$, which  gives a  RDM   corresponding to a separable mixed state. At   $t = n\pi/2$, with $n\in\mathbb{Z}$, the time-evolved state   falls into one of the separable states ($\ket{000}$ or $\ket{110}$), yielding zero concurrence $C=0$. Similarly, the RDM for the other adjacent pair, $\rho_{23}$, can be obtained to quantify the entanglement between sites 2 and 3. However, due to the symmetric coupling in our system, the behaviour of both $\rho_{12}$ and $\rho_{23}$ is equivalent, and we do not present $\rho_{23}$ explicitly. 

When it comes to the RDM for the two outer  QDs, $\rho_{13}$, which is essential for evaluating the next-nearest-neighbour  concurrence, we follow the same spirit as discussed above and in Subsec.\,\ref{concurrence3SKC}. Before proceeding with this analysis, we examine the structure of the time-evolved state in Eq.~(\ref{psit000}) with respect to the outer QDs (1 and 3). The time evolved state in Eq.~(\ref{psit000}) can be viewed as a superposition of two components: one involving a maximally entangled state of the form $\cos^2(\Delta t)\ket{00} - \sin^2(\Delta t)\ket{11}$ in sites 1 and 3, with the middle site in state $\ket{0}$; and another component of the form $-\frac{i}{2} \sin(2\Delta t)(\ket{01} + \ket{10})$  corresponding to the middle site in state $\ket{1}$. This decomposition clarifies when maximally entangled configurations such as $(\ket{00} \pm \ket{11})/\sqrt{2}$ or $(\ket{01} \pm \ket{10})/\sqrt{2}$ emerge during the dynamics.

Having provided some physical insights on the concurrence for the initial separable state   
$\ket{\psi_5(0)} = \ket{000}$ given by Eq.~(\ref{istate3}), in Fig.~\ref{C12_13commonEDensity}(a-b) we   present the nearest-neighbour concurrence $C_{12}$ as a function of time and  $\varepsilon_{i}=\varepsilon$ at $\Delta_{i}=\tau_{i}$, while in Fig.~\ref{C12_13commonEDensity}(b) as a function of time and $\varepsilon_{1}$. In Fig.~\ref{C12_13commonEDensity}(c), we show the  next-nearest-neighbour concurrence $C_{13}$ as a function of    time and  $\varepsilon_{i}=\varepsilon$, while in Fig.~\ref{C12_13commonEDensity}(d) we show the time evolution of $C_{12}$, $R_{p}$, and $E_{d}$ at the genuine sweet spot $\Delta_{i}=\tau_{i}$ and $\varepsilon_{i}$. Here, $C_{12}$ is obtained from Eq.\,(\ref{C12}), while $R_{p}$  and $E_{d}$ are obtained as discussed in Subsec.\,\ref{RpEd3SKC} for a target state given by Eq.\,(\ref{phi_state3}); see also Eq.\,(\ref{RP}) and Eq.\,(\ref{ED}). The first feature to note in Fig.~\ref{C12_13commonEDensity}(a) is that $C_{12}$ reveals the  generation of entanglement at all values of $\varepsilon$ quick after the evolution starts.   This behaviour arises because the initial wave function is localised in a single separable configuration, which begins to spread across the Hilbert space as the system evolves, leading to finite entanglement at all energies. Over time, the wave function settles into a structure determined by the coupling parameters, giving rise to distinct entanglement regimes indicated by the bright regions in Fig.~\ref{C12_13commonEDensity}(a). At the genuine sweet spot, $\Delta_{i}=\tau_{i}$ and $\varepsilon=0$, the concurrence $C_{12}$ exhibits a periodic behaviour with maxima occurring at $|\sin(4\Delta t)|/2 = 0.5$, which correspond  to   $t = \pm n\pi/(8\Delta)$, with $\Delta\equiv\Delta_{i}$.  Interestingly, by tuning the common onsite energy to a small finite value near $\varepsilon=0.5$, we see that $C_{12}$ takes larger values which is a clear signal of an enhanced  entanglement   between the first and second QDs in the two-site Kitaev chain, see extended yellow regions in Fig.~\ref{C12_13commonEDensity}(a). To get further insights on the entanglement at the genuine sweet spot, we compare the time evolution  of   $C_{12}$, $R_p$, and $E_d$ in Fig.~\ref{C12_13commonEDensity}(d) at the genuine sweet spot. Here, $C_{12}$ exhibits periodic behavior. We do not show $E_G$, which is monotonic with the concurrence  $C_{12}$.  In contrast, the return probability $R_p = \cos^4(\Delta t)$ is periodic, with revivals occurring at times $t = n\pi/\Delta$, where $n \in \mathbb{Z}$. The entanglement dynamics, given by $E_d = \cos^2(2\Delta t)/2$, is also periodic, but with half the period and half the amplitude of $R_p$. In the plots, $R_p$ is shown as a red dashed curve and $E_d$ as a green dashed curve. This happens because $E_d$ is defined to track the overlap between $\ket{\psi(t)}$ given by Eq.\,(\ref{psit000}) and the target state   $\ket{\phi_{2}}=(\ket{000} + \ket{101})/\sqrt{2}$ given in Eqs.\,(\ref{phi_state3}), namely, $E_{d}=|\bra{\phi_{2}}\ket{\psi(t)}|^{2}$, since their coefficients yield  $[\cos^2(\Delta t)-\sin^2(\Delta t)]/2 = \cos^2(2\Delta t)/2$ in the evolution. This indicates persistent maximal entanglement between the outer QDs, conditioned on the middle qubit being in the state $\ket{0}$.

The entanglement generation can be further enhanced by tuning  the onsite energies of one of the outer QDs. This is demonstrated in Fig.~\ref{C12_13commonEDensity}(b), where we show $C_{12}$ as a function of time and  $\varepsilon_{1}$ at $\varepsilon_{2,3}=0$ and $\Delta_{i}=\tau_{i}$. In this case, $C_{12}$ achieves particularly high values at large values of $\varepsilon_{1}$, which indicate the generation of maximal entanglement between the first and second QDs, see   yellow patches in Fig.~\ref{C12_13commonEDensity}(b). These regions of maximally entangled states persist for relatively longer durations and across a broader range of $\varepsilon_1$ compared to the case of a common finite onsite energy $\varepsilon$, as shown in Fig.~\ref{C12_13commonEDensity}(a).

In relation to the concurrence between next-nearest-neighbour QDs $C_{13}$,  Fig.~\ref{C12_13commonEDensity}(c) shows that it vanishes at the genuine sweet spot $\varepsilon_{i}\equiv\varepsilon=0$ at any time, indicating that  there is no entanglement between the first and third QDs at these onsite energies. This effect can be understood by noting that, to evaluate the concurrence  $C_{13}$, we compute the corresponding RDM $\rho_{13}$. At the sweet spot, this RDM [see Eq.~(\ref{concFormula})] satisfies the conditions $|\rho_{23}| = \sqrt{\rho_{11} \rho_{44}}$ and $|\rho_{14}| = \sqrt{\rho_{22} \rho_{33}}$, such that the differences used to quantify the concurrence vanish throughout the time evolution, resulting in  $C_{13} = 0$ and hence no    entanglement between first and third QDs as observed in Fig.~\ref{C12_13commonEDensity}(c) at $\varepsilon=0$. However, a finite onsite energy ($\varepsilon \neq 0$) introduces the necessary coherence that yields a finite positive difference in the expressions $|\rho_{23}| - \sqrt{\rho_{11} \rho_{44}}$ and $|\rho_{14}| - \sqrt{\rho_{22} \rho_{33}}$ for the concurrence in Eq.~(\ref{concFormula}), thereby generating a finite entanglement between the outer QDs of the three-site Kitaev chain [Fig.~\ref{C12_13commonEDensity}(c)].

Tuning the system away from the genuine sweet spot influences the generation of entanglement depending on the relative strengths of $\Delta_i$ and $\tau_i$. The corresponding results are presented in   Figs.~\ref{C12_13commonEDensity}(e-h) for $\Delta_{1}=\tau_{1,2}=1$ and $\Delta_{2}=0.5$. The nearest-neighbour concurrence $C_{12}$ as a function of time and $\varepsilon_{i}\equiv\varepsilon$ shows that strong entanglement continues to emerge around $\varepsilon = 0$, although its overall strength remains  around $C_{12} = 0.7$ and it is hence comparable to   the achieved value at the genuine sweet spot in Figs.~\ref{C12_13commonEDensity}(a).  Varying one of the onsite energies at the outer QDs, as shown in Fig.~\ref{C12_13commonEDensity}(f), can lead to the rapid formation of highly entangled states over a wide range of such onsite energies shortly after the evolution begins. However, this strong entanglement decreases as time progresses and subsequently revives at longer times, though within a narrower energy window, as shown by the smaller yellow patches in Fig.~\ref{C12_13commonEDensity}(f). To further understand entanglement at $\varepsilon=0$, in Fig.~\ref{C12_13commonEDensity}(h) we plot $C_{12}$, $R_{p}$, and $E_{d}$ as a function of time. Broadly speaking, all the entanglement measures develop an oscillatory time evolution with fast and slow frequency parts, which arise from the from having distinct values between $\Delta_{i}$ and $\tau_{i}$, namely, $\Delta_{1}\equiv\tau_{1,2}=1$ and $\Delta_{2}=0.5$. We do not show $E_G$ between first and second QDs, which follows the concurrence  $C_{12}$ monotonically. Meanwhile, $R_p$ takes a relatively long time to revive to the initial state, as seen in the orange dashed curve, due to the asymmetric couplings in the Hamiltonian.  The value of $E_d$ rises from $1/2$ because the initial state is $\ket{000}$, and it increases when contributions from the configuration $\ket{101}$ interfere constructively in phase, as shown in the green dashed curve. When it comes to the entanglement between outer QDs, we also find strong entanglement signatures. To show this, in Fig.~\ref{C12_13commonEDensity}(g) we plot the next-nearest-neighbour concurrence $C_{13}$ as a function of time and $\varepsilon$. In this case,   $C_{13}$  exhibits finite values over a broad range of onsite energies as time evolves. Unlike the genuine sweet spot ($\Delta_{i}=\tau_{i}$), $C_{13}$ interestingly acquires finite values even around $\varepsilon=0$ that weakens as time evolves, which implies a nonzero entanglement between the first and third QD at vanishing onsite energies.   

\begin{figure}[!t]
\centering
\includegraphics[width=0.98\linewidth]{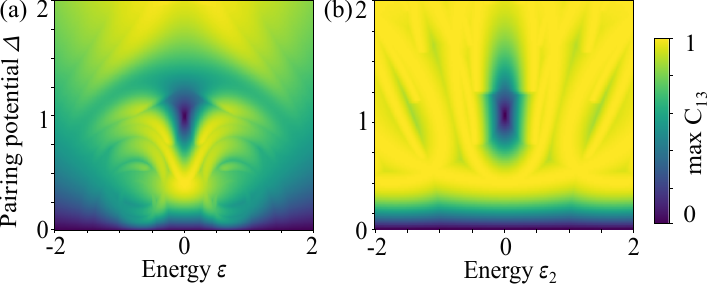}
\caption{(a) Maximum value of $C_{13}$ obtained over a sufficiently long time evolution to reach   maximal entanglement as a function of $\varepsilon_{i}\equiv\varepsilon$ and $\Delta_{i}\equiv\Delta$ for an initial separable state $\ket{000}$ and $\tau_{i} \equiv\tau= 1$ in a three-site Kitaev chain. Here, maximally entanglement is obtained for $t=7\, [1/\tau]$ so that it sets the interval over which  ${\rm max}$ $C_{13}$ is found.  (b) Same quantity as in (a) but plotted as a function of $\varepsilon_2$ and $\Delta_{i}=\Delta$ for $\varepsilon_{1,3}= 0$.}
\label{maxC13Density}
\end{figure}

To gain deeper insights into the behaviour of next-nearest-neighbour entanglement, we now calculate the maximum value of the concurrence $C_{13}$, denoted as ${\rm max}\,C_{13}$, obtained over a sufficiently long time evolution to reach the maximal entanglement from the initially separable state $\ket{000}$ in the three-Kitaev chain. In Fig.~\ref{maxC13Density}, we show ${\rm max}\,C_{13}$ as a function of $\varepsilon_{i}=\varepsilon$ and $\Delta_{i}=\Delta$. We observe that  $C_{13}\rightarrow0$ at very small values of $\Delta$ and at the genuine sweet spot  ($\tau =\Delta and \varepsilon = 0$), the latter consistent with Fig.~\ref{C12_13commonEDensity}(c). Moreover, $C_{13}$ remains finite throughout the rest of the parameter space, reaching even values $C_{13}\rightarrow1$ that signal maximal entanglement between outer QDs. As a consequence of the behavior of $C_{13}$, we can conclude that the absence of  entanglement between the outer QDs at sizeable $\Delta$ is an indicator of the genuine sweet spot. This behaviour is further confirmed in Fig.~\ref{maxC13Density}(b), where $C_{13}$ is plotted as a function of the onsite energy of the middle QD, $\varepsilon_2$.   In this case, $C_{13}$ reaches its maximum entanglement value over a broader parameter range (yellow regions), except at  $\Delta=\tau$ and $\varepsilon_{2}=0$, which corresponds to the genuine sweet spot. This behaviour closely resembles the maximally entangled states observed in the dynamics of the effective two-site Kitaev chain, as discussed in the energy analysis below Eq.~(\ref{even3sweetspot}).

In summary, the three-site Kitaev chain exhibits richer entanglement dynamics than its two-site counterpart: detuning from the sweet spot can enhance both nearest-neighbour and long-range entanglement, and depending on the parity of the initial states, multipartite GHZ- or W-type states can be dynamically generated. This highlights the potential of minimal three-site chains as versatile platforms for the controlled generation of both bipartite and multipartite entanglement, depending on the specific application.

So far, in quantifying bipartite entanglement within the three-site Kitaev chain, we have not considered a maximally entangled initial state analogous to that used in Eq.~(\ref{iState}) for the two-site Kitaev chain. This is because its extension to the three-site chain, introduced in Eq.~(\ref{istate3}), is a superposition of even- and odd-parity configurations, which evolve independently during the dynamics. Moreover, this state corresponds to a genuinely multipartite entangled state.  As a result, its dynamics not only reveals bipartite entanglement as discussed earlier in this section but also provides a platform for exploring multipartite entanglement, which we address in the next section.

%%%%%%%%%%%%%%%%%%%
%%%%           Section V        %%%%%
%%%%%%%%%%%%%%%%%%%

\section{Multipartite entanglement in a three-site Kitaev chain}
\label{section5}
Going beyond bipartite entanglement between pairs of QDs, in this section we quantify multipartite entanglement by employing the   geometric measure of entanglement (GME) as described in Subsec.~\ref{gm3sites}.  As discussed in that subsection, multipartite entanglement primarily falls into two distinct classes: GHZ- and W-type entanglement \cite{PhysRevA.68.042307}. To study the generation of GHZ-type entanglement, one must consider an initial state that spans both the even- and odd-parity sectors of the Hamiltonian. Qualifying this criterion, the GHZ state introduced in Eq.~(\ref{istate3}) serves as a suitable initial state, as it extends the two-site maximally entangled state and includes superpositions of configurations from both parity sectors of the Hamiltonian. In contrast, multipartite entanglement in W-type states can be investigated by considering initial states lying in either parity sector of the Hamiltonian, since W-type states are defined within a specific parity subspace. Therefore, the quantification of such an  entanglement is independent of the specific initial state chosen in Eq.~(\ref{istate3}).

\subsection{Geometric measure of multipartite entanglement in reference to the GHZ and W states}
To quantify multipartite entanglement, we begin by analysing the dynamics of the composite parity sector initialised in the GHZ state defined by Eq.~(\ref{istate3}). As discussed earlier, this initial state is well-suited for capturing multipartite entanglement and enables the simultaneous investigation of both GHZ- and W-type entangled state dynamics. The time-evolved state of the system takes the  general form given by
\begin{equation}
\label{t_aribrary3qubitstate}
\begin{split}
\ket{\psi(t)} &= \alpha_1(t)\ket{000} + \beta_1(t)\ket{011} \\
&+ \gamma_1(t)\ket{101} + \delta_1(t)\ket{110} \\
                  &+  \alpha_2(t)\ket{001} + \beta_2(t)\ket{010} \\
                  &+ \gamma_2(t)\ket{100} + \delta_2(t)\ket{111} \,,
\end{split}
\end{equation}
where the coefficients $\alpha_i(t),\, \beta_i(t),\, \gamma_i(t),\, \delta_i(t)$ with $i = 1,2$ are time-dependent and determined by the Hamiltonian parameters. This state provides clearer understanding into the composite dynamics of the even- and odd-parity sectors of the Hamiltonian, which results from the time evolution of the configurations $\ket{000}$ and $\ket{111}$ of the GHZ state in Eq.~(\ref{istate3}). This composite evolution is  necessary for the generation of multipartite entanglement in the GHZ state.  In contrast, to analyse entanglement in the W-type state, we can restrict the analysis to either parity sector of Eq.~(\ref{t_aribrary3qubitstate}), which can also be obtained by independently evolving either of the initial configurations $\ket{000}$ or $\ket{111}$, depending on the parity of the W-type state under consideration. However, the W state we consider is  $W_{e}$ and   belongs to the even parity sector of the Hamiltonian, see   the paragraph below Eq.\,(\ref{sep3state}). This implies that, to assess entanglement in the W state,   the coefficients with $i=1$ in Eq.\,(\ref{t_aribrary3qubitstate}) are playing a role.

To compute the GME, we use Eq.\,(\ref{Eg2site}) which requires to calculate the magnitude of the overlap between a separable three-qubit state $\ket{\phi}$ [Eq.\,(\ref{sep3state})]  and the time-evolved state $\ket{\psi(t)}$, namely, $\Lambda=|\bra{\phi}\ket{\psi(t)}$, where $\ket{\phi}$ and  
$\ket{\psi(t)}$ are given by Eq.\,(\ref{sep3state}) and Eq.~(\ref{t_aribrary3qubitstate}), respectively.
These overlaps are obtained following the same spirit of Subsec.~\ref{gm3sites} but with the arbitrary time-evolved state given in Eq.~(\ref{t_aribrary3qubitstate}). Here, in the overlap $\Lambda$, the separable state $\ket{\phi}$ is chosen to be symmetric, whereas the dynamically generated states are, in general, asymmetric. Consequently, the symmetric separable ansatz employed here provides an analytic upper bound to the GME when the dynamical state moves away from the space of symmetrically entangled states \cite{PhysRevA.68.042307}, and is therefore used as a qualitative indicator of multipartite entanglement generation.

Since now we want to address the GME in the time evolved state $\ket{\psi(t)}$ given by Eq.\,(\ref{t_aribrary3qubitstate}), to compare the achievement of GHZ and W maximally entangled states, we have to be careful about the  contributions from specific configurations coming from GHZ and W states. For instance, when looking 
if $\ket{\psi(t)}$ reaches the GHZ state, we only keep the coefficients  from the $\ket{000}$ and $\ket{111}$ since their superposition defines the  GHZ state; this overlap is defined as $\Lambda^{\rm GHZ}=|\bra{\phi}\ket{\psi(t)}|_{\rm GHZ}$,  where the lower script  GHZ indicates that we collect coefficients corresponding to $\ket{000}$ and $\ket{111}$ configurations. A similar process is carried out to identify if $\ket{\psi(t)}$ achieves the W state, and the overlap in this case is defined as $\Lambda^{\rm W}=|\bra{\phi}\ket{\psi(t)}|_{\rm W}$, although all the configurations are present since we are considering an imperfect W state, see paragraph below Eq.\,(\ref{sep3state}). Then, we find that the magnitude of the W and GHZ overlaps  
\begin{equation}
\label{LamMax3even}
\begin{split}
\Lambda^{\rm GHZ} &=| \alpha_1(t)\cos^3\theta + \delta_2(t) \sin^3\theta|\,,\\
\Lambda^{\rm W} &=| \alpha_1(t)\cos^3\theta \\
&+ \left[ \beta_1(t) + \gamma_1(t) + \delta_1(t) \right] \sin^2\theta \cos\theta|\,.
\end{split}
\end{equation}
By maximising the above  overlaps numerically over $\theta\in[0,\pi]$, we compute the GME with respect to both separable states as $E^{\nu}_{\rm G}=1-[\Lambda_{\rm max}^{\nu}]^{2}$, where $\Lambda_{max}^{\nu}={\rm max}_{\theta}[\Lambda^{\rm \nu}]$ and $\nu={\rm W, GHZ}$.

 Before going further, we notice that $\Lambda^{\rm W}$ in Eqs.\,(\ref{LamMax3even}) acquires a contribution due to $\alpha_1(t)$ as a result of the presence of   $\ket{000}$ in the state dynamics; this  is different to $\Lambda^{\rm W}$ in Eqs.~(\ref{overlapf}) which does not have a contribution from $\ket{000}$. This contribution arises not only from the chosen initial state, but also from the structure of the system’s eigenstates [see Eq.~(\ref{even3sweetspot})], which retain $\ket{000}$ contributions even when evolving from other configurations within the even-parity sector of the Hamiltonian. This leads to the formation of an imperfect W-type state of the form $\alpha \ket{000} + \beta \ket{W_e}$, as discussed in Subsec.~\ref{gm3sites}. 

\begin{figure}[!t]
\includegraphics[width=1\linewidth]{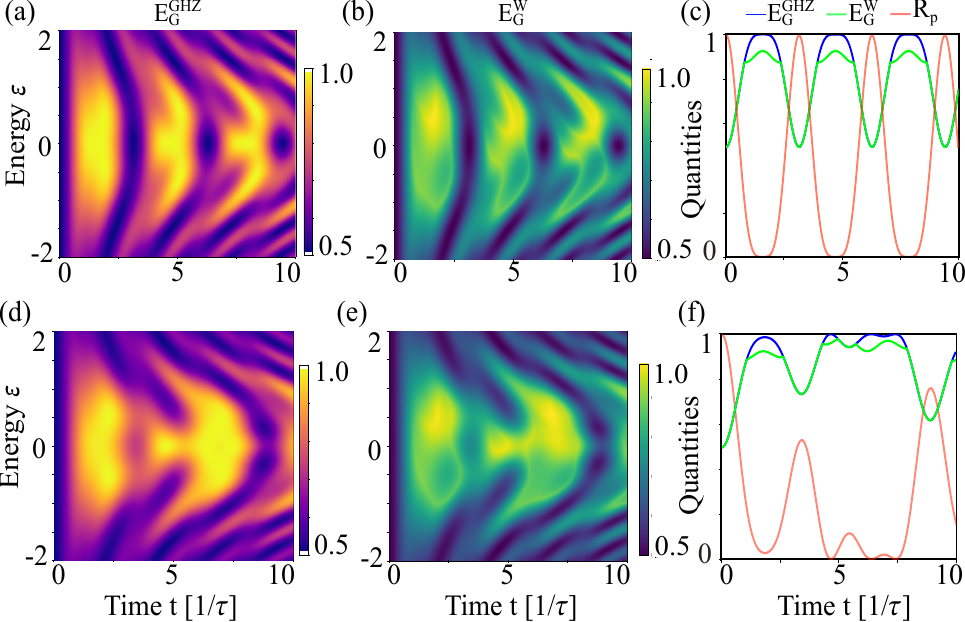}
 \caption{Multipartite GME as a function of  time $t$ and onsite energy $\varepsilon_{i}=\varepsilon$   at the sweet spot $\Delta_{i}=\tau_{i}$ (a,b) and away from the sweet spot $\Delta_{i}\neq\tau_{i}$ (c,d)  in a three-site Kitaev chain with an initial GHZ   state $(\ket{000}+\ket{111})/\sqrt{2}$. (a,b) GME with respect to the GHZ state ($E^{\rm GHZ}_{\rm G}$) and with respect to the imperfect W state ($E^{\rm W}_{\rm G}$). (c)  Time evolution of $E^{\rm GHZ}_{\rm G}$, $E^{\rm W}_{\rm G}$, and $R_p$ at $\varepsilon = 0$. (d-f) Same as in (a-f), but for $\Delta_{i}\neq\tau_{i}$.  Parameters: (a-c) $\Delta_{i}=\tau_{i}=1$, while (d-f) $\Delta_{1}=\tau_{1,2}$, $\Delta_{2}=0.5$.}
 \label{fig:ge_ghz_Wtype}
\end{figure}

\subsubsection{At and away from the sweet spot}
To identify the GME with respect to the GHZ and W states, in    Fig.~\ref{fig:ge_ghz_Wtype} we plot $E^{\rm GHZ}_{\rm G}$ and $E^{\rm W}_{\rm G}$ obtained from Eqs.\,(\ref{LamMax3even})  as a function of time $t$ and onsite energy $\varepsilon_{i}=\varepsilon$ at the genuine sweet spot ($\Delta_{i}=\tau_{i}$) and away from it ($\Delta_{i}\neq\tau_{i}$). Moreover,  for comparison, in   Fig.~\ref{fig:ge_ghz_Wtype}(c,f) we also show the time evolution of $E^{\rm GHZ}_{\rm G}$ and $E^{\rm W}_{\rm G}$ along with the the return probability $R_p$ for the GHZ initial state given in Eq.~(\ref{istate3}). %An analogous comparison with the perfect W state is not feasible in this system due to the persistent presence of the $\ket{000}$ amplitude, which renders our W state imperfect as discussed earlier. 

The first feature one observes at the genuine sweet spot in Fig.~\ref{fig:ge_ghz_Wtype}(a,b) is that the GME $E^{\rm GHZ}_{\rm G}$ and $E^{\rm W}_{\rm G}$ develop bright and dark regions that oscillate between $1/2$ and $1$. We remind that that reaching maximally GHZ state entanglement means $E^{\rm GHZ}_{\rm G}=1/2$, while $E^{\rm W}_{\rm G}=5/9$  for  W state entanglement; away from these values, the system realizes a separable state (yellow regions). In this regard, $E^{\rm GHZ}_{\rm G}=1/2$ for $\varepsilon=0$, a value that evolves in time following an almost periodic profile  indicating the easiness  to achieve  a GHZ state, see Fig.~\ref{fig:ge_ghz_Wtype}(a) and also blue curve in Fig.~\ref{fig:ge_ghz_Wtype}(c). This behaviour is particularly pronounced over a broad range of onsite energies and at short timescales.  Since the dynamics of $\ket{000}$ and $\ket{111}$ evolves similarly in Eq.~(\ref{t_aribrary3qubitstate}), the GHZ state undergoes periodic revivals at times $t = n\pi$ (with $n$ an integer), when the amplitudes of both configurations reach their maximum within their respective parity sectors. As a result, $E^{\rm GHZ}_{\rm G}\rightarrow1/2$,  indicating a genuinely multipartite entangled state, as highlighted by the darker regions in Fig.~\ref{fig:ge_ghz_Wtype}(a) and blue curve in Fig.~\ref{fig:ge_ghz_Wtype}(c). When $\varepsilon$ is tuned to finite values away from zero, i.e., in the upper or lower regions relative to the $\varepsilon = 0$ line, the entanglement becomes less robust  and the dynamics lose their periodic structure on shorter timescales but it is still possible to identify multipartite  GHZ entanglement.  

In the case of  $E^{\rm W}_{\rm G}$ in Fig.~\ref{fig:ge_ghz_Wtype}(a), the dynamics also facilitates the generation of strong multipartite entanglement. This entanglement manifests as an imperfect W-type state, where the $\ket{000}$ component remains present. Notably, the generated entanglement via $E^{\rm W}_{\rm G}$  exceeds the value of $5/9$ corresponding to the perfect W state \cite{PhysRevA.68.042307, PhysRevLett.92.087902}, see discussion below Eq.~(\ref{overlapf}). At  $\varepsilon=0$, $E^{\rm W}_{\rm G}$   periodically oscillates in time, depicted by  the dark regions in Fig.~\ref{fig:ge_ghz_Wtype}(b) and green curve in Fig.~\ref{fig:ge_ghz_Wtype}(c); here, {these dark blue values reach $1/2$ which are close to the value $5/9$ of the perfect W state GME.} Thus, the lighter blue regions that appear between the dark blue and yellow areas indicate zones of stronger entanglement. However, this entanglement is less stable than the GHZ-type entanglement found near $\epsilon=0$  and are  characterized by broader regions of either dark blue or yellow [Fig.~\ref{fig:ge_ghz_Wtype}(a)]. In this context, the higher values of $E^{\rm W}_{\rm G}$ approaching unity represent a separable configuration, see yellow regions in  Fig.~\ref{fig:ge_ghz_Wtype}(b) and green curve in  Fig.~\ref{fig:ge_ghz_Wtype}(c).  Still, strong entanglement emerges robustly at finite onsite energy values, corresponding to   lighter blue regions in [Fig.~\ref{fig:ge_ghz_Wtype}(b)].

Further understanding on the behavior of $E^{\rm GHZ}_{\rm G}$ and $E^{\rm W}_{\rm G}$ at the genuine sweet spot $\Delta=\tau$ and $\varepsilon=0$ can be obtained by comparing their time evolution with the  return probability $R_p$ in Fig.~\ref{fig:ge_ghz_Wtype}(c). Initially, both $E^{\rm GHZ}_{\rm G}$ and $E^{\rm W}_{\rm G}$ start at 0.5 with $R_p = 1$, reflecting the choice of the GHZ state as the initial state. As the system evolves, the initial state spreads across the Hilbert space, and the corresponding entanglement measures for GHZ- and W-type states evolve accordingly. Since the initial state is maximally entangled with respect to the GHZ state, the overlap function in Eq.~(\ref{LamMax3even}) with the GHZ state decreases over time, leading to an increase in  $E^{\rm GHZ}_{\rm G}$, which reaches its maximum value of 1 when $\Lambda_{max}^{\rm GHZ}$ in Eq.~(\ref{LamMax3even}) vanishes. This condition is satisfied at $t = (n + 1)\pi / 2$, where $n$ is an integer, and corresponds to $\alpha_1 = \cos^2(\Delta t) = 0$, as seen from Eq.~(\ref{t_aribrary3qubitstate}). The first zero occurs at $t_1 = \pi/2 \approx 1.57$. At these times, the contributions from the $\ket{000}$ and $\ket{111}$ components vanish [see Eqs.~(\ref{psit000}) and (\ref{psitodd})], causing the return probability $R_p$ to drop to zero. Along with $\ket{000}$, two additional coefficients in Eq.~(\ref{t_aribrary3qubitstate}) also approach zero, as confirmed in Eq.~(\ref{psit000}). This drives the system towards a separable state, as the eigenvalues in Eq.~(\ref{LamMax3even}) decrease. Consequently, the value of  $E^{\rm W}_{\rm G}$, associated with an imperfect W-type state, also increases in a manner similar to that of the GHZ entanglement. However, during the dynamics, $E^{\rm W}_{\rm G}$ can never reach unity, in contrast to  $E^{\rm GHZ}_{\rm G}$. This is because the separable state into which the system evolves retains configurations present in the imperfect W state, resulting in a finite $\Lambda_{max}^{\rm W}$ in Eq.~(\ref{LamMax3even}).

Moving away from the sweet spot, i.e., for $\Delta_{i} \neq \tau_{i}$, the evolution of all quantities becomes aperiodic over sufficiently long timescales, shown in the bottom panels of Fig.~\ref{fig:ge_ghz_Wtype}, where $\Delta_1=\tau_{1,2}=1$ and $\Delta_2=0.5$. This leads to stronger multipartite entanglement for finite $\varepsilon$ as can be noted by the proliferation of more darker regions in Figs.~\ref{fig:ge_ghz_Wtype}(d).  Contrary to the genuine sweet spot case,  finite values of $\varepsilon$ do not lead to stronger entanglement during a short time evolution, as indicated by the dominating yellow regions in  $E^{\rm GHZ}_{\rm G}$ and $E^{\rm W}_{\rm G}$, see Figs.~\ref{fig:ge_ghz_Wtype}(d) and \ref{fig:ge_ghz_Wtype}(e), respectively.  To compare $E^{\rm GHZ}_{\rm G}$ and $E^{\rm W}_{\rm G}$ with the return probability   $R_p$, in Fig.~\ref{fig:ge_ghz_Wtype}(f) we present their time evolution at $\varepsilon=0$ which uncovers an irregular behavior with time. In this regime, a revival of $E^{\rm GHZ}_{\rm G}$ is not observed at short times. The first revival occurs much later, around $t \approx 33[\tau^{-1}]$, which lies outside the time window shown in the plots.

To close this section, we emphasize that the geometric measure of entanglement explored for the three-site Kitaev chain unveils the emergence of multipartite entanglement, with robust values both at and away from the genuine sweet spot. Moreover, the properties of the geometric measure of entanglement, together with other dynamical quantities, shows that a three-site Kitaev chain can achieve maximally entangled GHZ-type states as well as imperfect W-type states.

\section{Conclusions}
\label{SectionVI}
In conclusion, we have investigated dynamics of bipartite and multipartite entanglement in two- and three-site Kitaev chains,  systems that have already been realized using quantum dots coupled by superconductor-semiconductor hybrids.  In particular, we   characterized    bipartite and   multipartite entanglement at and away from the sweet spot  by employing the concurrence and the geometric measure of entanglement, which we also complemented   by  using dynamical observables such as return probability and entanglement dynamics.

In the two-site Kitaev chain, we have shown that bipartite maximally entangled states can be generated regardless of the choice of initial state, with the entanglement behavior sensitively dependent on the system parameters. Specifically, we found that, for an initially unentangled state, the most stable maximally entangled states emerge when the system is tuned away from the sweet spot.   At the sweet spot, however, the system oscillates between separable and maximally entangled states.  Moreover, we found that by varying one of the onsite energies, the dynamics of the entanglement measures exhibit tunable   valleys between successive maximally entangled states, thus offering a way to control the stability of maximally entangled states. For an initially maximally entangled state, we showed that the entanglement measures exhibit a similar behaviour, although, as expected due to the choice of the initial state, high entanglement is more likely to occur.

In the case of three-site Kitaev chain, we demonstrated that the system exhibits both bipartite and multipartite entanglement  possessing qualitatively distinct features. At the genuine sweet spot, bipartite entanglement between nearest-neighbour quantum dots oscillates between separable and partially entangled states, reaching up to half of the maximum concurrence.  In contrast, we found that  the next-nearest-neighbour concurrence between the outer quantum dots vanishes at the genuine sweet spot but attains a finite value when the system is away from it. We   showed that both types of bipartite entanglement can be enhanced by detuning any pair of quantum dots from the sweet spot either  by varying the onsite energy of a quantum dot, the pair potential, or the hopping parameter. Interestingly, in the three-site Kitaev chain,   the geometric measure of entanglement revealed the emergence of   multipartite entanglement, with robust values both at and away from the sweet spot. In this case,  we discovered that the dynamics of the geometric measure  can achieve maximally entangled states of the GHZ type, as well as imperfect W-type states.

Altogether, by combining concurrence, geometric entanglement, entanglement dynamics, and return probability within a unified dynamical framework, our analysis of two- and three-site Kitaev chains establishes a systematic approach to generating and characterizing highly entangled states with both bipartite and multipartite structure. Given the current experimental advances in two- \cite{Dvir2023,bordin2023tunable,bordin2024crossed,zatelli2024robust,ten_Haaf2024,van_Loo2026} and three-site Kitaev chains \cite{ten_Haaf2025,Bordin2025,Bordin2026}, our results also provide a concrete foundation for assessing entanglement resources in minimal Kitaev chain and their potential relevance for quantum information and computation.\cite{r75t-jv32,van_Loo2026,PRXQuantum.5.010323,PhysRevB.109.075101}. Our results also raise important questions, such as what is the role of environment effects for generating bipartite and multipartite quantum correlations in minimal Kitaev chains; this would require an open system description and the analysis of other measures, such as quantum discord, constitutes an interesting direction for future research.

 %%%%%%%%%%%%%%%%%%%%%%%%%%%%%%%
%                        ACKNOWLEDGMENTS                               %
%%%%%%%%%%%%%%%%%%%%%%%%%%%%%%%

\section{Acknowledgements}
We thank M. Benito, M. Leijnse, and E.  Sj\"{o}qvist for insightful discussions. We acknowledge financial support from the Swedish Research Council  (Vetenskapsr\aa det Grant No.~2021-04121) and from  the Carl Trygger’s Foundation (Grant No. 22: 2093).

%%%%%%%%%%%%%%%%%%%%%%%%%%%%%%%
%                                  APPENDIX                                  %
%%%%%%%%%%%%%%%%%%%%%%%%%%%%%%%

\onecolumngrid

\appendix

\section{Odd-sector dynamics in a three-site Kitaev chain}
\label{oddDyn3sites}
For completeness and for comparison with the Hamiltonian's even-parity analysis presented in Subsec.~\ref{evenDynsep3KC}, we now examine the dynamics of the odd-parity initial state $\ket{111}$. To this end, we first express the Hamiltonian in the odd-parity subspace [see Eq.~(\ref{Odd3HMat})], spanned by the basis states $\ket{001}$, $\ket{010}$, $\ket{100}$, and $\ket{111}$, at the sweet spot ($\Delta = \tau$, $\varepsilon = 0$):
\begin{equation}
\label{oddHSweetspot}
\mathcal{H}_{3\mathrm{KC}}^{o'}=  \left(\begin{array}{cccc}
0 &  \Delta &  0 & \Delta 	\\
\Delta  & 0 & \Delta & 0 \\
0 & \Delta & 0  &  \Delta 	\\
\Delta & 0  & \Delta & 0
\end{array}\right)\,,
\end{equation}
This matrix has the same structure as the one used to represent the Hamiltonian in the even sector at the sweet spot, which can be derived from Eq.~(\ref{Even3HMat}).  As a result, its eigenvalues and eigenvectors can be directly obtained by mapping the even-parity configurations to their corresponding odd-parity counterparts from Eq.~(\ref{even3sweetspot}). The eigenvalues remains the same $E_0^i = \{ -2\Delta, 2\Delta, 0, 0 \}$, and the corresponding eigenvectors are given by
\begin{equation}
\label{odd3sweetspot}
\begin{aligned}
&
\ket{E_{o}^{1(2)}} = \frac{1}{2}\bigl[\mp\ket{001} + \ket{010} \mp \ket{100} + \ket{111}\bigr], \\[6pt]
&
\ket{E_{0}^3 } = \frac{1}{\sqrt{2}}\bigl[-\ket{001} + \ket{100}\bigr], \\[6pt]
&
\ket{E_{0}^4} = \frac{1}{\sqrt{2}}\bigl[-\ket{010} + \ket{111}\bigr].
\end{aligned}
\end{equation}
At first, comparing the above eigenstates of the odd-parity sector with those of the even-parity sector [Eq.~(\ref{even3sweetspot})], one can conclude that the dynamics of $\ket{000}$ and $\ket{001}$ are equivalent. This arises because both sectors share identical eigenvalues and similar eigenvector structures, and the time evolution is governed by the first three eigenstates in each case. However, a closer inspection reveals that the first and fourth eigenstates in Eq.~(\ref{odd3sweetspot}) differ from their even-sector counterparts in Eq.~(\ref{even3sweetspot}) by a global phase of $-1$ with respect to the configurations $\ket{000}$ and $\ket{111}$, respectively. This equivalence leads to identical dynamical behavior for $\ket{000}$ and $\ket{111}$ within their respective parity sectors. Explicitly, the dynamics in odd sector with initial state $\ket{111}$ can be written as
\begin{equation}
\label{psitodd}
    \ket{\psi(t)} = \cos^2(\Delta t)\ket{111} - \sin^2(\Delta t)\ket{010}  - \frac{i}{2}\sin(2\Delta t)[\ket{001}+ \ket{100}] \,.
\end{equation}
As a result, the time evolution of the full system, initialised in the GHZ state defined in Eq.~(\ref{istate3}), yields two analogous expressions at the sweet spot for the odd and even parity sectors, which can be written in the form of Eq.~(\ref{t_aribrary3qubitstate}).

\section{Geometric measure of entanglement in a two-site Kitaev chain}
\label{GME2site}

In this Appendix, we show the geometric measure of entanglement $E_G$ at the sweet spot for system parameters exactly used in Fig.\,\ref{All00compare} to compare with the other entanglement measures. As we have shown in the final expressions for all entanglement measures in Eq.\,(\ref{SepQuantities}), $E_G$ comes out to be a monotonic function of concurrence, its results follows the similar patterns of the concurrence shown in Fig.\,\ref{All00compare}.

\begin{figure*}[ht]
  \centering
  \includegraphics[scale=0.98]{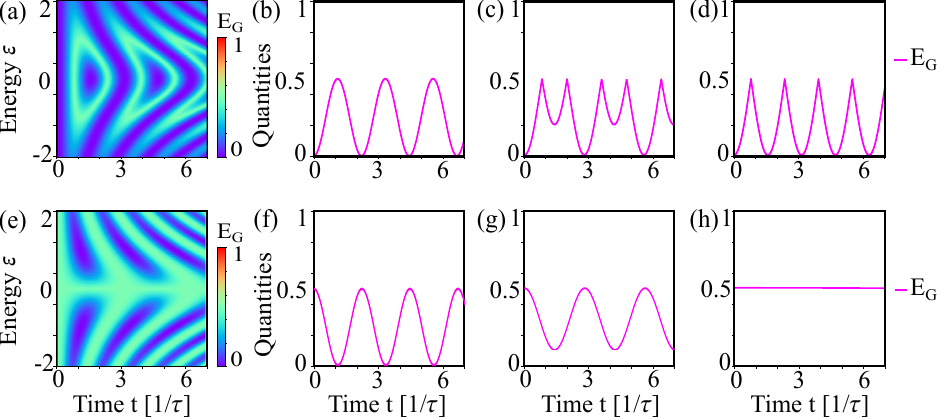}
 \caption{Geometric measure of entanglement $E_G$ at the sweet spot ($\tau = \Delta$) for an initially separable state $\ket{00}$ (a-d) and for an initially maximally entangled state $(\ket{00} + \ket{11})/\sqrt{2}$ (e-h). (a) $E_G$ as  functions of time $t$ and onsite energy $\varepsilon_{i}=\varepsilon$ at the sweet spot ($\tau = \Delta$).  (b)  Time evolution of $E_G$ at $\varepsilon = 1$. (c,d) Same as in (b), but for  $\varepsilon_1 = 0$, $\varepsilon_2 = 1$ and 
   $\varepsilon_1 = \varepsilon_2 = 0$, respectively. Bottom panel:  (e-h) Same as in (a-d), but for a maximally entangled initial state $(\ket{00} + \ket{11})/\sqrt{2}$.   Parameters: $\Delta=\tau=1$. } 
 \label{All00compare_Eg}
\end{figure*}

To provide a comparative visualization with other quantities shown in Fig.~\ref{All00compare}(a-d), we present the time-evolution of $E_G$, given by Eqs.\,(\ref{SepQuantities}) at distinct onsite energies $\varepsilon_{i}$ and at $\Delta=\tau$ on the same pattern. For $E_G$, the dark blue regions indicate unentangled states, whereas the bright light-blue regions correspond to maximally entangled states; see Fig.~\ref{All00compare_Eg}(a). Similar to concurrence (C), it exhibits a periodic generation of entanglement, ranging from completely unentangled ($E_G = 0$) to maximally entangled states ($E_G = 0.5$), and shows qualitatively similar behaviour to the concurrence shown in Fig.~\ref{All00compare}(a) of the main text. Following $C$, the dynamics of $E_{G}$ exhibit similar behaviour when the sign of $\varepsilon$ is reversed. At the sweet spot $\varepsilon = 0$, $E_G$ exhibits the dynamics of separable and maximally entangled states in a periodic manner over time [Fig.~\ref{All00compare_Eg}(a)]. Increasing the value of $\varepsilon$ from 0 to 0.5 nearly doubles the frequency of achieving maximally entangled states [Fig.~\ref{All00compare_Eg}(a)], as evident from the increased number of light-blue patches formed in the plot of $E_{G}$.  Following Eq.\, (\ref{Sweet00Quantities}) in the main text, the periodicity of $E_G$, as for the concurrence, is given by $t = (2n + 1)\pi / (2\Delta)$, where $n$ is an integer.

The dynamical characteristics of $E_G$ for distinct onsite energies at $\Delta = \tau = 1$ are shown in Fig.~\ref{All00compare_Eg}(b-d). Specifically, the time evolution of $E_G$ is presented in Fig.~\ref{All00compare_Eg}(b) for $\varepsilon_{1,2} = 1$, in Fig.~\ref{All00compare_Eg}(c) for $\varepsilon_{1} =0,  \varepsilon_{2} = 1$, and in Fig.~\ref{All00compare_Eg}(d) for $\varepsilon_{1,2} = 0$. As discussed earlier, for all parameter sets $E_G$ varies monotonically with the concurrence $C$. However, for onsite energies $\varepsilon_{1} =0,  \varepsilon_{2} = 1$ and $\varepsilon_{1,2} = 0$, unlike $C$ shown in Fig.~\ref{All00compare}(d) and Fig.~\ref{All00compare}(e), respectively, $E_G$ exhibits relatively sharp peaks at its maxima, as shown in Figs.~\ref{All00compare_Eg}(c) and \ref{All00compare_Eg}(d). This behaviour can be attributed to the definition of $E_G$, for which the derivative can diverge, as indicated by Eqs.\,(\ref{SepQuantities}).

For the sweet-spot regime ($\varepsilon_{1,2} =  0$, $\Delta = \tau = 1$), Fig.~\ref{All00compare_Eg}(d) illustrates a perfectly periodic $E_G$ that alternates between zero and its maximum. Following $C$, the unentangled states occur whenever $\cos(\Delta t) = 0$ or $\sin(\Delta t) = 0$ in Eqs.\,(\ref{Sweet00Quantities}), yielding a period of $\pi/2$. Consequently, $E_G$ reaches its maximum at $t = (2n + 1)\pi/4$ in the genuine sweet spot of a two-site Kitaev chain.

Moving to the dynamics of initially maximally entangled states, as well as other cases for the three-site Kitaev chain, $E_G$ remains monotonic with the concurrence $C$. We do not show these results explicitly, as they do not exhibit qualitative differences and are omitted to avoid repetition.

\twocolumngrid

\bibliography{biblio}

\end{document}